\documentclass[preprintnumbers,amsmath,amssymb,superscriptaddress,twocolumn,showpacs]{revtex4-1}
\usepackage{graphicx}
\usepackage{dcolumn}
\usepackage{bm}
\usepackage{natbib}
\usepackage{physics}
\usepackage[caption=false]{subfig}

\newcommand{\be}{\begin{equation}}
\newcommand{\ee}{\end{equation}}
\newcommand{\bea}{\begin{eqnarray}}
\newcommand{\eea}{\end{eqnarray}}

\begin{document}


\title{Magnetic-Torque Enhanced by Tunable Dipolar interactions}
 
\author{C. Pellet-Mary$^1$, P. Huillery$^1$, M. Perdriat$^1$, G. H\'etet} 

\affiliation{Laboratoire De Physique de l'\'Ecole Normale Sup\'erieure, \'Ecole Normale Sup\'erieure, PSL Research University, CNRS, Sorbonne Universit\'e, Universit\'e de Paris , 24 rue Lhomond, 75231 Paris Cedex 05, France.}

\begin{abstract}
We use tunable dipolar-interactions between the spins of nitrogen-vacancy (NV) centers in diamond to rotate  
a diamond crystal.
Specifically, we employ cross-relaxation between the electronic spin of pairs of NV centers in a trapped diamond to enhance the anisotropic NV paramagnetism and thus to increase the associated spin torque.
Our observations open a path towards the use of mechanical oscillators to detect paramagnetic defects that lack optical transitions, to investigation of angular momentum conservation in spin relaxation processes and to novel means of cooling the motion of mechanical oscillators.
\end{abstract}

\maketitle

Controlling the motion of macroscopic oscillators at ultra low motional temperatures has been the subject of intense research over the past decades.  In this direction, opto-mechanical systems, where the motion of micro-objects is strongly coupled to laser light, have had tremendous success \cite{Aspelmeyer}. 
Similar interaction schemes were propounded in order to strongly couple long-lived atomic spins, such as the electronic spin of nitrogen-vacancy (NV) centers in diamond, to mechanical oscillators in the quantum regime \cite{Treutlein2014, Rabl, Lee_2017}.  At the single spin level, this achievement would offer the formidable prospect of transferring the inherent quantum nature of electronic spins to the oscillators, with foreseeable far-reaching implications in quantum sensing and tests of quantum mechanics \cite{Bose, Wan, Ma}.

Most efforts using single NV centers are presently hampered by their low coupling strengths to the motion, which are currently far below typical spin decoherence rates \cite{Kolkowitz, Gieseler, DelordNat, Arcizet}.   
One solution to counteract this issue is to work with large ensembles of spins \cite{DelordNat}. This approach does not lend itself easily to observing non-linear spin-mechanical effects, but may offer a more favorable path towards ground state spin-cooling \cite{Rabl} and would enable the observation of many-body effects mediated by the motion \cite{Wei, Ma}. 

\begin{figure}[!ht]
  \centering \scalebox{0.1}{\includegraphics{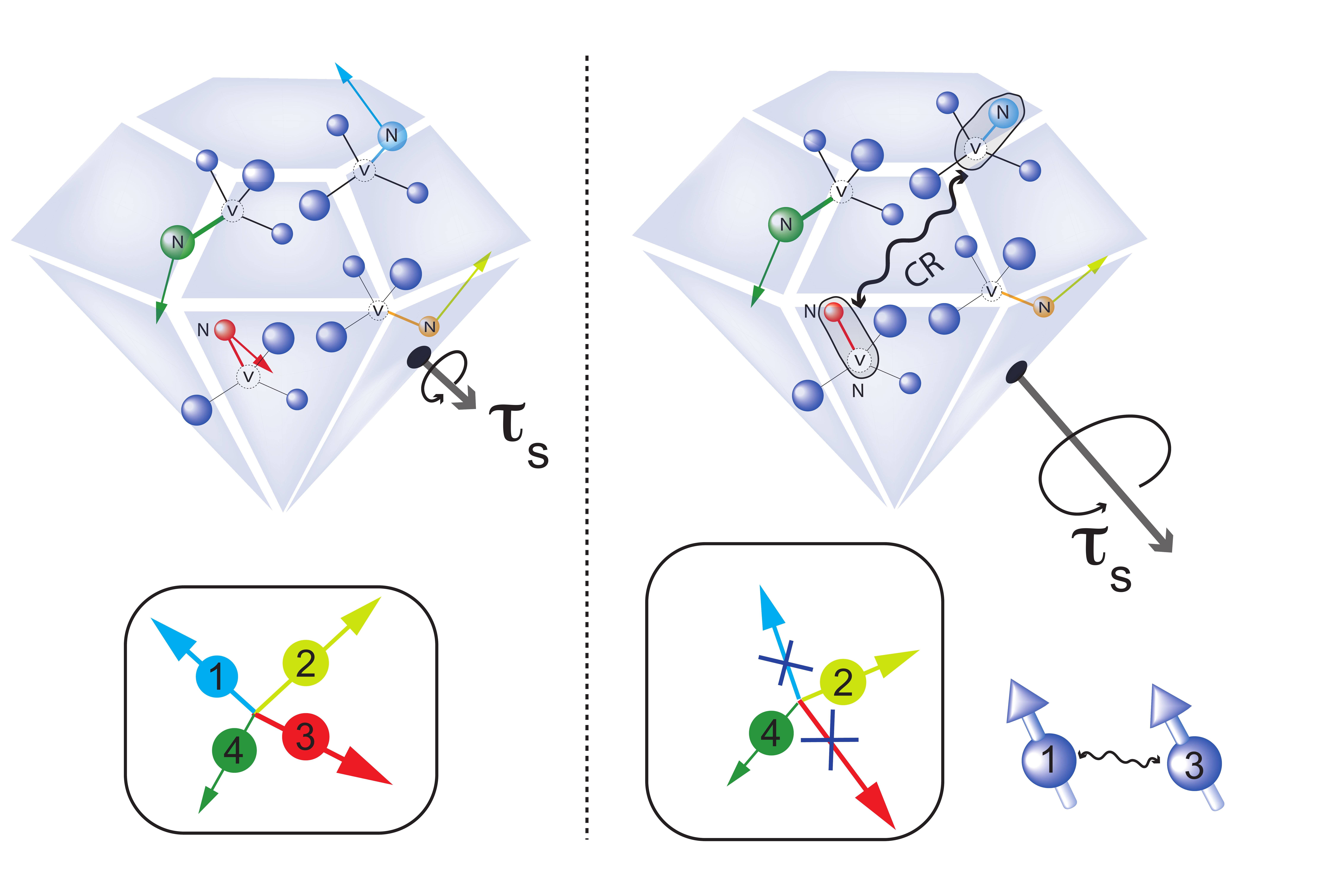}}
  \caption{General principle of the resonant dipole-dipole enhanced mechanical rotation. The four possible directions of the nitrogen-vacancy centers in the diamond are shown in the left/right panels together with their spin-torque contributions (arrows of the corresponding colors). Left panel: the (quasi)-rotational invariance gives a small total spin torque $\tau_s$.
Right panel: A magnetic field (not shown) is tuned so that the spin class 1 and 3 point to the same direction. Cross-relaxation (CR) between these two classes of NV centers occurs, altering the rotational symmetry and increasing $\tau_s$. 
  }\label{principle}
\end{figure}

However, although the spin-mechanical coupling strength is predicted to increase linearly with the number of spins, this scaling-law is modified when the mean distance between the atomic defects is of the order of 10~nm because of dipolar interactions. Dipolar interactions can significantly enrich the physics at play and have for instance been employed in the optical domain to increase the coupling of electron dipoles to mechanical motion, akin to super-radiant processes \cite{Bachelard, PANAT, Venkatesh, Juan}. Closely packed NV centers may also show similar cooperative effects. Further, the coupling strength can be tuned resonantly amongst different NV orientations \cite{van_oort_cross-relaxation_1989}, offering prospects for studying the interplay between dipolar interactions and motional degrees of freedom in a controlled fashion. Increasing the density of NV centers also means that they can couple to other spins in the diamond \cite{armstrong_nvnv_2010, Alfasi, Epstein, Hall} and even transfer their polarization \cite{WangBajaj}.
Angular momentum exchange in such cross-relaxation processes could result in a rotation of the crystal, 
as in the Einstein-de-Haas effect, and even enable controlling mechanical oscillators in the quantum regime \cite{Zangara}.  

\begin{figure}[!ht]
  \centering \scalebox{0.08}{\includegraphics{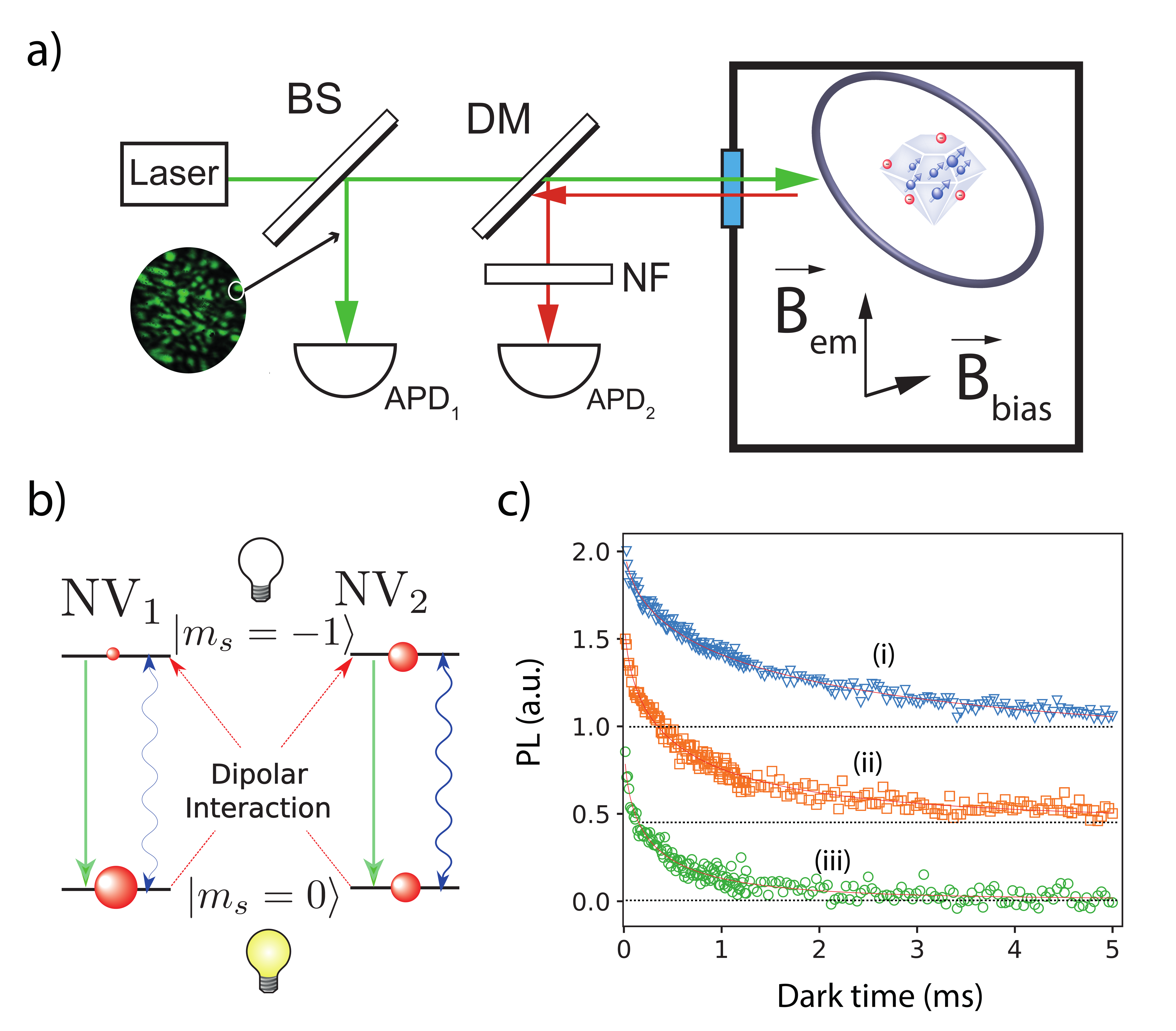}}
  \caption{Schematics of the experiment. A micro-diamond is levitating in a ring Paul trap enclosed in a vacuum chamber. A green laser is used both to polarize the NV centers in the levitating diamond and to detect the angular motion. Part of the speckle pattern formed in the image plane is sent onto APD$_1$ after passing through a beam splitter (BS). The photoluminescence from the NV centers is collected on APD$_2$ after filtering out the green laser light by a dichroic mirror (DM) and a notch filter (NF). a) Sketch showing the NV-NV cross-relaxation process. Green arrows represent the optical pumping to the brighter $\ket{m_s=0}$ state. The two curvy blue arrows with different thicknesses represent short/long longitudinal relaxation of NV$_2$/NV$_1$. Red circles represent the population in each state and red dashed arrows represent the resonant dipole-dipole interaction between the two NV centers. c) Measurements of the longitudinal relaxation from a single NV class when, i) it is not resonant with any other classes ($T_1=1.61$~ms), ii) when it is resonant with another class ($T_1=490~\mu$s) and iii) when it is resonant with the three other classes ($T_1=220~\mu$s). The three traces have been offset for clarity.
  }\label{setup}
\end{figure}

Here, we employ resonant dipolar interactions to rotate a micro-mechanical oscillator. 
Specifically, we use NV centers inside a diamond that is levitating in a Paul trap that is similar to the one used in \cite{DelordPRL} and use resonant cross-relaxation (CR) between them to observe a spin-torque coming from the NV paramagnetism. The key mechanism is depicted in Fig.~\ref{principle}. 
As depicted in the left panel, NV centers are found in four different orientations in the diamond crystalline structure. 
As will be shown next, in the presence of an external transverse magnetic field, NV centers acquire a magnetization.
Due to quasi-rotational invariance of the problem, although each NV class could exert a significant magnetic torque to the diamond, the total spin-torque $\tau_s$ is reduced, and the resulting paramagnetic susceptibility is of the order of the diamagnetism from the electrons of the carbon atoms.
However by tuning an external magnetic field, resonant dipole-dipole interactions between the spin of NV centers of different orientations is enhanced which, in turn, increases the paramagnetism.

When the spin transition of NV centers become resonant, the polarization of the different orientations can be exchanged through cross-relaxation \cite{Abragam}. The conditions on the magnetic field for CR to occur are described in Sec. I of the Supplementary Material (SM)\cite{SM_CR_meca}. \nocite{DelordPRL} \nocite{delordPhD} \nocite{DelordNat} \nocite{qutip1} \nocite{qutip2} \nocite{van_oort_cross-relaxation_1989} \nocite{choi_depolarization_2017}
The right panel of Fig.~\ref{principle}, shows a CR mechanism that partly removes the contribution from two classes of NV centers (labelled 1 and 3 in Fig.~\ref{principle}), which breaks the four-spin rotational invariance. The total spin torque $\tau_s$ can then be large enough to rotate the diamond.

%

\begin{figure}[!ht]
  \centering \scalebox{0.24}{\includegraphics{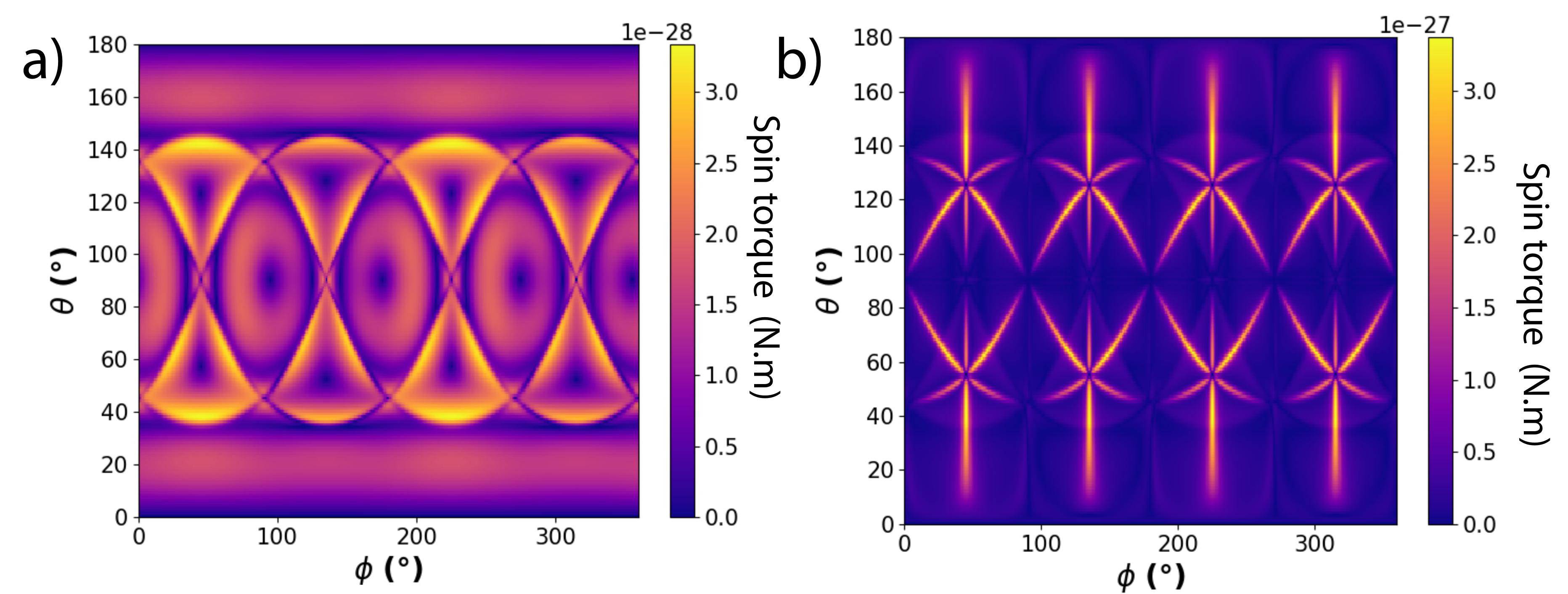}}
  \caption{Numerical simulations of the spin-torques on a diamond containing one NV center per orientation as a function of $\theta$ and $\phi$, the polar and azimuthal angle with respect to the [100] direction. a) and b) show the torque with and without cross-relaxation between NV centers respectively. Notice the different torque scales.}
  \label{Numerics}
\end{figure}

It was shown in \cite{choi_depolarization_2017} that in highly doped diamond samples, a few fast-decaying NV centers, so called {\it fluctuators} can depolarize an ensemble of NV centers through dipolar interaction. Fig.~\ref{setup}-b) depicts the dipolar interaction between two NV centers. In this example, the electronic spin of NV$_1$ is polarized in the ground state via the green laser, whereas NV$_2$ is a fluctuator, which has a shorter relaxation time $T_1$ than the polarisation time.
The spins will exchange magnetic quanta through flip-flop processes resulting in a depolarization of NV$_1$. This was shown to reduce the average $T_1$ of the ensemble from the phonon-limited $T_1$ ($\approx$ ms) to a few hundreds of micro-seconds \cite{Jarmola} and to lower the total photoluminescence \cite{van_oort_cross-relaxation_1989, armstrong_nvnv_2010, jarmola_longitudinal_2015, akhmedzhanov_microwave-free_2017, akhmedzhanov_magnetometry_2019, holliday_optical_1989, mrozek_longitudinal_2015, choi_depolarization_2017} in bulk materials. The origin of the fast-decaying NV centers was attributed to the presence of charge tunneling amongst closely packed NV centers \cite{choi_depolarization_2017}. The NV centers that undergo tunneling with other impurities (possibly with the substitutional nitrogen defect \cite{manson_nv_2018}) have a largely reduced longitudinal spin lifetime $T_1$.

Such a process has not been studied in detail with nano- or micro-particles to the best of our knowledge. Smaller diamond particles in fact tend to suffer from extra parasitic surface effects such as spin depolarization due to interaction with paramagnetic dangling bonds on the surface \cite{Tetienne}, or enhanced charge transfer between the NV$^0$ and NV$^-$ charge states \cite{Dhomkar} so it is essential to verify that it can be observed with micro-particles. 
We start by searching for CR using micro-diamonds that are physically attached to the trap, by employing a fixed bias magnetic field $||\bf B_{\rm bias}||\approx$100 G and by tuning another magnetic field $\bf B_{\rm em}$ at some angle with respect to $\bf B_{\rm bias}$ using an electromagnet (see Fig.~\ref{setup}-a)). The change in orientation of the total magnetic field can be visualized in Sec. I of the SM \cite{SM_CR_meca}.

The photoluminescence from the NV centers is detected using standard confocal microscopy. 
At specific magnetic field directions with respect to the crystalline axes, degeneracy between the spin of NV centers can be reached \cite{van_oort_cross-relaxation_1989}.
We measured the $T_1$ time in these conditions by applying a green laser that polarizes the NV centers and measure the photoluminescence at a later time.
Such a measurement can be significantly impacted by recharging of NV centers in the dark \cite{choi_depolarization_2017, mrozek_longitudinal_2015, giri_selective_2019, giri_coupled_2018}.
In order to accurately measure the $T_1$ and remove the changing PL due to the recharging effects, we use the sequence presented in Sec. III of the SM \cite{SM_CR_meca}, where a microwave pulse is applied or not prior to spin relaxation. The PL signals acquired in the two different measurements are then subtracted and shown for different degeneracy configurations in Figure \ref{setup}-c).
In the absence of degeneracy, we observe a stretched-exponentially decaying profile \cite{choi_depolarization_2017}, from which we extract a $T_1=1.61$~ms, already shorter than the phonon limited lifetime in dilute bulk materials \cite{Tetienne}. This lifetime is even further reduced when more orientations are brought to resonance. This hints towards the role played by dipolar interactions, which are enhanced when more classes of NV centers are resonant \cite{van_oort_cross-relaxation_1989, choi_depolarization_2017}. 


The main goal the present study is to demonstrate mechanical action of such dipolar induced relaxations when diamonds are levitating in the Paul trap.
One major extra ingredient for this is the induced magnetization of the NV centers when they are polarized in the ground state, which has thus far not been directly observed.
Let us consider first the dependence of the ground state energy of a single spin as a function of the angle between a magnetic field and the NV axis.
The Hamiltonian for one NV orientation with quantization axis $z'$ in the particle frame reads 
\begin{equation}\hat{H}_{\rm NV}=\hbar D \hat{S}_{z'}^2+ \hbar \gamma_e \bf B  \cdot \bf\hat S,
\end{equation}
where $\bf\hat S$ is the spin-vector, $D=(2\pi)2.87$ GHz the zero-field splitting and $\bf B$ is the external magnetic field.
Under the condition $\gamma ||\bm B|| \ll D$, assuming an NV center in the $(x,z)$ plane and a B field along $z$, $\hat H_{B}=  \hbar \gamma_e {\bf B}  \cdot {\bf \hat S}=\hbar\gamma_e B  ( \hat S_{x'} \sin\theta + \hat S_{z'} \cos\theta)$ can be treated as a perturbation to the anisotropic part $\hbar D \hat{S}_{z'}^2$ of the Hamiltonian. Here, $\theta$ is the angle between the magnetic field and the body-fixed NV center axis.
The energy $\epsilon_g$ of the ground state perturbed by the B field is then
\begin{equation} \epsilon_g=\sum_{m_s=\pm 1}\frac{ |\bra{0} \hat H_B \ket{\pm 1}|^2}{-\epsilon^0_{\pm 1}}=-\hbar\frac{(\gamma_e B_\perp)^2 }{D},
\end{equation}
where $B_\perp=B \sin\theta$.
\begin{figure*}[!ht]
  \centering \scalebox{0.45}{\includegraphics{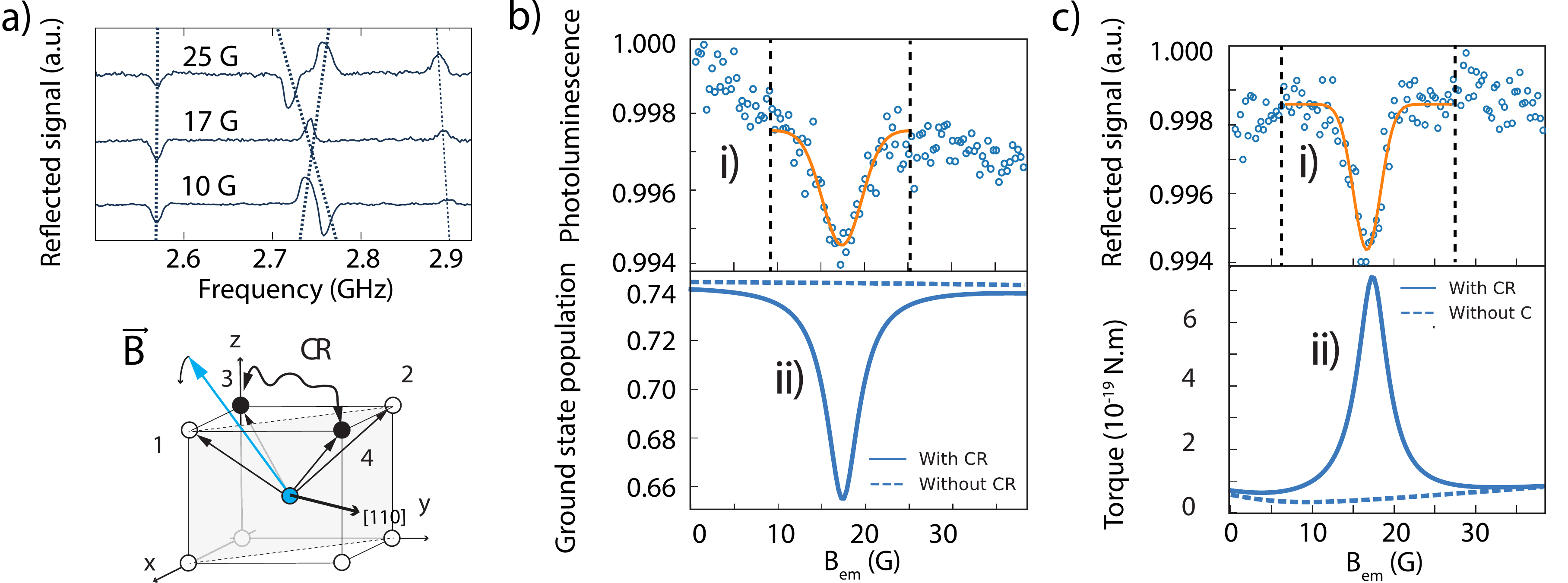}}
  \caption{a) Top :  Signal reflected off the diamond surface as a function of microwave frequency for three different magnetic field values. Bottom : sketch showing the crossing of a crystal plane when the magnetic field angle is tuned.
b) PL detection as a function of B$_{\rm em}$ across a dipole-dipole resonance. i) experimental data with gaussian fit, ii) simulation of the population in $\ket{m_s=0}$ state, taking into account (plain lines) or not (dashed lines) cross-relaxation.
c) Angular detection as a function of B$_{em}$ across a resonance. i) experimental data with gaussian fit, ii) simulation of the magnetic torque applied to the diamond.
  }
  \label{data}
\end{figure*}
A direct use of the Hellmann-Feynman theorem can give the torque in the ground state. We find that 
\begin{equation}\tau_s=- \frac{\partial \epsilon_g}{\partial \theta}=\hbar  \frac{(\gamma_e B)^2}{D} \sin 2\theta.\label{Eq1}
\end{equation}
A proof of the applicability of this theorem in the presence of dissipation is presented in  Sec. IV of the SM \cite{SM_CR_meca}. 
At an angle $\theta=\pi/4$, where the torque is maximized and at a B field of 100 G, we obtain  $\tau_s\approx 2 \times 10^{-27}$N.m.
Taking into account the whole NV level structure, we then find $\tau_s\approx 10^{-18}$ N.m, using $10^9$ spins polarized in the ground state. Taking a librational confinement frequency of the diamond in the Paul trap to be around $\omega_\theta/(2\pi)\approx 1$ kHz, we obtain an spin-torque induced angular displacement 
of $\tau/I_y \omega_\theta^2\approx$1 mrad, which can be measured with a high signal-to-noise ratio in our set-up \cite{DelordNat}. Here $I_y\approx 10^{-22}$ kg.m$^2$ is the moment of inertia of the particle around the $y$ axis.

As already hinted to however, the contributions from the other NV classes must also be taken into account (see Fig. 1).
Fig. \ref{Numerics} presents the result of numerical calculations of the torque coming from the four classes of NV centers, assuming only one NV per orientation here.   
Fig. \ref{Numerics}-a) shows the torque magnitude as a function of $\theta$ and $\phi$ without taking into account CR. The torque from each of the four classes appear clearly from the symmetry. Their different contributions however sum up to give a maximum torque of around $10^{-28}$ N.m, which is 20 times smaller than the torque that can be obtained for a single class. The quasi-rotational invariance of the problem thus hinders the diamond paramagnetism. When two classes of NV center are resonant however, the induced cross-relaxation partly breaks this rotational invariance. 
Fig. \ref{Numerics}-b) shows the same plot, but including CR. Details on the model can be found in sec.VI of the SM \cite{SM_CR_meca}.
Here we use numbers that are deduced from the experimental observations of the CR-induced change of the $T_1$ in Fig. 2 b).
One can see that a new pattern with larger spin-torque is superimposed to the previous map. These larger values coincide with crossings of the crystal planes where NV degeneracies occur. 
At these coordinates, one recovers the torque estimation of Eq.~\ref{Eq1}, found for a single class, which would then imply a spin-torque that overcomes the Paul trap confinement. 

To observe the effect of such resonant dipolar interactions on the motion, we use similar parameters and magnetic field arrangement than when the diamonds were not levitating. The diamond crystalline direction with respect to the magnetic field direction is characterized by recording Mechanically-Detected-Magnetic-Resonances (MDMR) \cite{DelordNat} similar to in Magnetic Resonance Force Microscopy (MRFM) \cite{Rugar}. 
The angle motion is detected by collecting the back-reflected green light from the diamond interface (see Fig. 2-a), separated from the excitation light using a beam splitter as a microwave drives the spin to the $m_s=-1$ state.
Fig. \ref{data}-a) shows MDMR detection of spin-resonances for three different $\bf B_{\rm em}$ amplitudes. 
At 10 and 25~G, one can observe 4 peaks in the spectrum that demonstrate microwave-induced torque on the diamond from the 4 classes of NV centers.
At 17~G however, two classes merge at a microwave frequency of 2.75 GHz. This is where we expect to observe CR. 

A detailed analysis developed in Sec.I of the SM \cite{SM_CR_meca} suggests that since we observe a single degeneracy at 17~G, the magnetic field crosses a plane that is perpendicular to the $[110]$ direction, as shown in Fig. \ref{data}-a).
Fig. \ref{data}-b) shows the photoluminescence as a function of $\bf B_{\rm em}$ both experimentally (trace i) and numerically (trace ii).
As expected, the PL decreases across the degeneracies at around the same magnetic field value. 
Fig. \ref{data}-c), trace i) is a measurement of the diamond angular position acquired simultaneously to the PL. Trace ii) is the corresponding calculation.   
A pronounced variation of the reflected signal is also observed, demonstrating the close correspondence between degeneracy and diamond rotation, and the enhanced spin-torque as the dipolar interactions between the spins increase. Note that, as opposed to the PL detection which always shows dips in spectra, the laser signal coming from the particle surface can increase or decrease on resonance, depending on how the speckle is aligned to the fiber. This explains the differing shapes of the signals in the experiments and the simulations. 
Fitting trace c-i) by a Gaussian curve, we deduce a width that is similar to the PL width of trace b-i) (2.1~G and 2.8~G respectively). This gives a width of 9 (resp. 12) MHz  comparable to the inhomogeneous broadening of the sample.
Similar experiments were realized on different particles under different degeneracies.
In Sec. V of the SM, we present results taken under a two-fold degeneracy. 

Let us conclude by mentioning the applications offered by dipole-dipole induced mechanical rotation.
First, when performed under vacuum \cite{vacuumESR}, this effect can be employed to control the temperature and stiffness of mechanical oscillators in the absence of microwave.
For cooling, a delay between the spin and Paul trap torques \cite{Aspelmeyer,DelordNat} will be introduced by tuning the polarizing laser power to reach a depolarizing rate ($\approx 10$ kHz) of the order of the trapping frequency.
At a magnetic field value corresponding to a negative detuning from the CR feature, the NV fluctuator will depolarize a pair of spins and let the two other NV classes apply a torque until the preivous pair re-polarizes, extracting energy from the angular motion during each cooling cycle \cite{Braginski}.  

Conversely, the CR-induced torque can be viewed as a novel spectroscopic technic for sensing dipolar interactions between NV centers and spins that cannot be polarized optically.  Using a magnetic field oriented close to the diamond [111] direction would for instance enable detection of dark paramagnetic species that do not have a zero-field splitting \cite{WangBajaj}. The method may open a path towards the, otherwise difficult, experimental investigations of angular momentum conservation during relaxation processes in crystals, as proposed in \cite{Zangara}.

Last, and more prospectively, one could consider the presented technique to lay the grounds for bottom-up investigations of magnetism. 
The detailed microscopic origin of magnetism depends strongly on the material and spins have relaxation times that are typically very short (typically picoseconds), making microscopic investigations a complicated task.  In our employed paramagnetic sample, both the interaction between spins and their relaxation strength can be tuned on $\mu$s timescales. 
In addition, the present sensitivity $\eta = \sqrt{4 k T \gamma I_y }\approx 10^{-20}$~N.m.$/\sqrt{\rm Hz}$ of the torque sensor can be largely improved by going under high vacuum and using smaller particle sizes. Here $\gamma\approx (2\pi)1$ kHz is the damping rate of the angular motion due to collisions with gas molecules. Under $10^{-2}$~mbar and using 1~$\mu$m diameter particles would already give $\eta\approx 10^{-24}$~N.m.$/\sqrt{\rm Hz}$, approaching state of the art sensitivities \cite{Kim, Jonghoon} and thus opening a path towards using NV centers in levitating diamonds for emulating magnetism at the hundreds of Bohr magneton level \cite{carlin}.

\section*{acknowledgements}
GH acknowledges SIRTEQ for funding.

\newpage

\begin{widetext}

\vspace{0.2in}
{\Large \hspace{2.13in}\textsc{Supplementary Material} }\\

\section{NV$^-$ center Theory}
\subsection{NV spin hamiltonian}
The hamiltonian of the electronic spin of the negatively charged nitrogen-vacancy center in its ground state can be written as :
\begin{equation*}
  \hat{\mathcal{H}}_s=\hbar D \hat S_{z'}^2 + \hbar \gamma_e \textbf{B}\cdot\hat{\textbf{S}},
  \end{equation*} 
where $D = (2\pi) 2.87$ GHz is the crystal field splitting originating from spin-spin interactions, and $\gamma_e = 28 $GHz/T is the electron gyromagnetic ratio. 
The \textbf{z'} axis in the $\hat S_{z'}$ operator here is the axis formed by the nitrogen atom and the vacancy in the body fixed frame.
We neglect contributions from the strain and local electric field in the hamiltonian since we are working with magnetic fields on the order of 10 mT, which induce splittings larger than the splitting of the zero-field ESR line ($\approx 20$ MHz). We also neglect the hyperfine interaction with the nuclear spin of the $^{14}$N atom since we are working with ensembles with typical inhomogeneous broadening of 5~MHz. 


\subsection{Diamond crystalline axes and degeneracy conditions}
\begin{figure}[!ht]
  \centering \scalebox{0.45}{\includegraphics{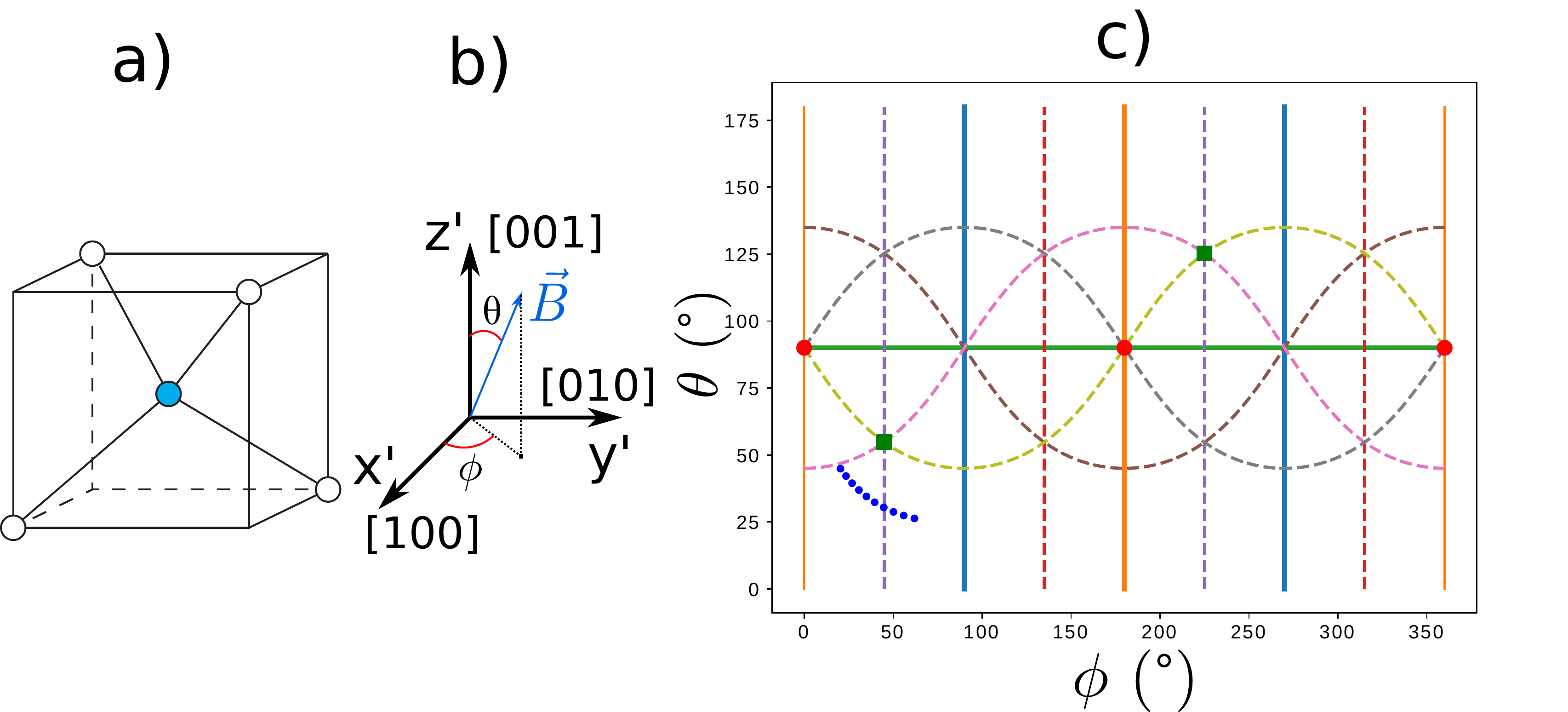}}
  \caption{a) Representation of the four crystalline axes of the diamond b) Representation of the magnetic field in the diamond crystalline basis c) Representation of the crystalline planes in the $(\theta , \phi)$ basis. The $\{ 110 \}$ family of planes are shown using dashed lines and the $\{ 110 \}$ family using plain lines. The [100] direction is marked by red circles, the [111] direction by green squares. The magnetic field path in the experiment of Fig. 3 of the main text is shown with blue dots.}
	\label{cristallo}
\end{figure}
There are four possible crystalline axes for the N-V direction (so-called ``classes" of NV) in the diamond. They are depicted in Fig. \ref{cristallo} b) and correspond to the crystalline directions [$111$], [$1\bar 1 \bar 1$], [$\bar 1 1 \bar 1$] and [$\bar 1 \bar 1 1$]. 

The magnetic field direction is represented in Fig. \ref{cristallo} a), where the polar and azimuthal angles $\theta$ and $\phi$ are defined with respect to the \textbf{z'} ([$001$]) direction (we denote with ' the axes in the diamond frame). For some orientations of the magnetic field, the projection of the magnetic field on two or more NV axes will be identical, and therefore the energy level of the corresponding classes will be the same. These degeneracies are represented in Fig \ref{cristallo} c), where the dashed lines are the locii of the $\{ 110 \}$ family of planes (plane normal to the [110] direction and all other equivalent directions, making 6 planes in total). When the magnetic field belongs to these planes, we observe a degeneracy between two classes of NVs, as can be seen in the Fig.\ref{scan_dege} or in Fig. 4 of the main paper.

The plain lines are the locii of the $\{ 100 \}$ family of planes (3 planes in total). When the magnetic field lies in these planes, all classes are co-resonant, as can be seen in Fig.\ref{scan_dege} or in Fig.\ref{CR_22}.
The red circles correspond to the [100] directions, for which the four classes of NV are degenerate. The green squares correspond to the [111] direction where one class is aligned with the magnetic field, and the three others are degenerate. Finally the blue dots correspond to the path followed by the magnetic field in the experiment presented in Fig 3 of the main text, where we can see that a plane from the $\{ 110 \}$ family is being crossed.

\section{Depolarization induced by NV-NV cross-relaxation}

Our diamonds are supplied by the company Adamas, which produces diamonds with a concentration of NV centers in the 3-4 ppm range. 
As explained in the main text, when the density of NV$^-$ centers in the sample is large enough (typically for concentrations higher than 1 ppm), the ensemble of NV spins will lose some of its polarization through dipolar coupling between the NV centers. This phenomenon is at the heart of the mechanism that allows us to exalt the magnetic susceptibility of our diamond through dipolar interaction, and it has already been observed independently by many groups in bulk diamond \cite{Jarmola_SI} \cite{mrozek_longitudinal_2015_SI} \cite{choi_depolarization_2017_SI} \cite{akhmedzhanov_microwave-free_2017_SI} \cite{giri_coupled_2018_SI}.

In particular, \cite{choi_depolarization_2017_SI} proposes a model based on "fluctuators" : a subgroup of NV centers with a very short lifetime (possibly due to their electron tunneling in and out of the NV site) can act as a source of classical noise with a central frequency given by the transition frequencies of the NV$^-$ spin Hamiltonian. One prediction of this model is that the modified lifetime of the ensemble of NV centers should have a stretch exponential profile (of the form $e^{-\sqrt{\frac{t}{T_1}}}$). We do observe this scaling law experimentally.

\subsection{Stretch exponential profile of the lifetimes}
\begin{figure}
\includegraphics[scale=0.7]{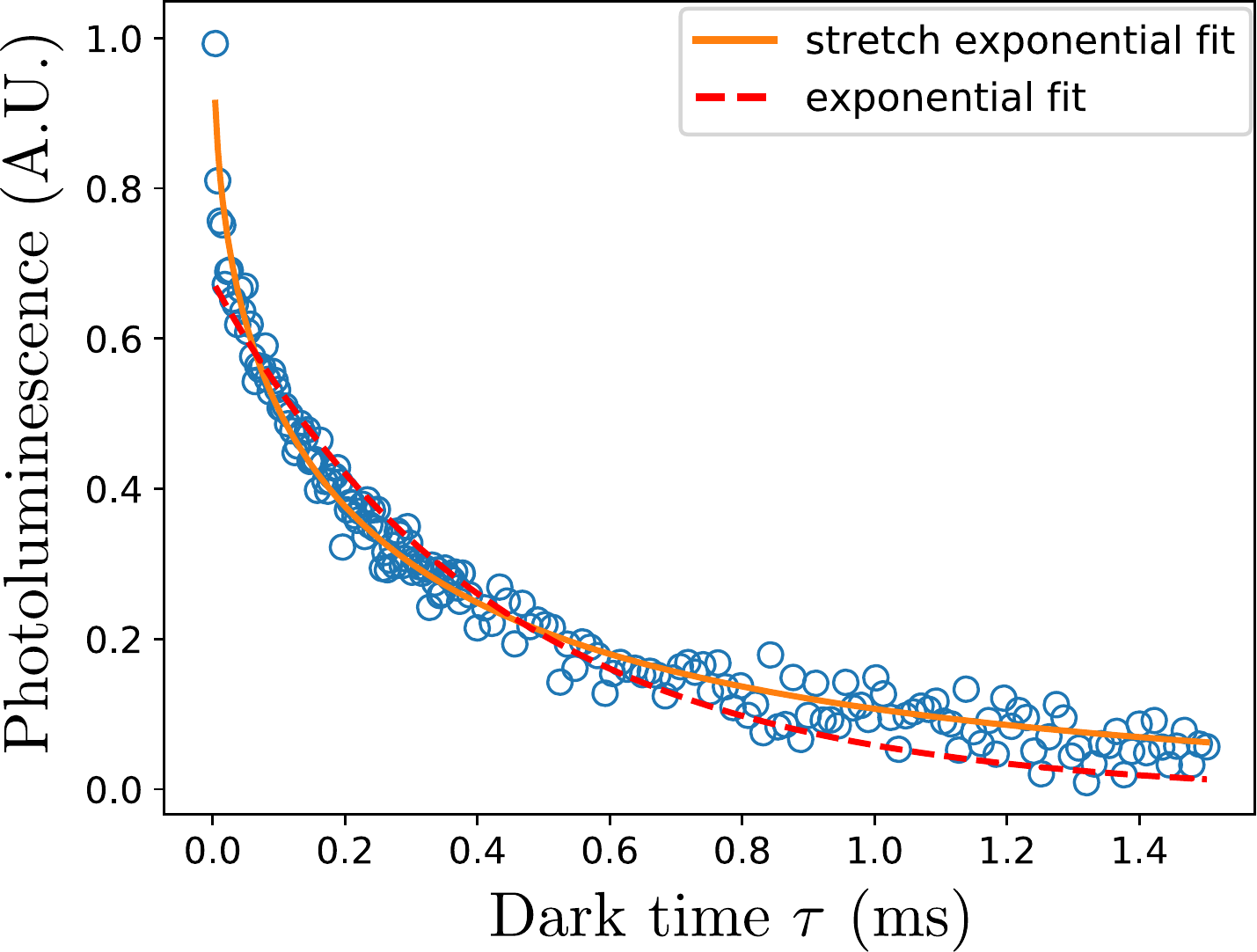}
\caption{Lifetime measurement when all four NV classes are degenerate. Plain orange line correspond to a stretch-exponential fit and dashed red line to a simple exponential fit}
	\label{stretch}
\end{figure}
In the theory developed in \cite{choi_depolarization_2017_SI}, the stretch exponential profile arises from the inhomogeneity of the distance from each NV centers to the closest fluctuators. We write $\rho_{00}^s(t)$, the population in the $\ket{m_s=0}$ state for each NV centers evolving in the dark. This population follows a law of the form $\rho_{00}^s(t) \propto \exp(-\gamma t)$ where $\gamma$ is the individual depolarization rate of the spin; then, assuming an homogeneous spatial distribution of fluctuators, the authors of \cite{choi_depolarization_2017_SI} show that the distribution in $\gamma$ should follow a law of the form $$\rho(\gamma) \approx \frac{e^{-\frac{1}{4 \gamma T_1}}}{\sqrt{4 \pi \gamma^3 T_1}} $$ where $\rho(\gamma)$ is the density of probability of $\gamma$.

Averaging then over all NV centers gives the stretch exponential profile observed from the ensemble :
$$ \rho_{00}^e(t) \propto \int_0^{+\infty} \rho(\gamma) e^{-\gamma t} d\gamma = e^{-\sqrt{\frac{t}{T_1}}}, $$
where $ \rho_{00}^e(t)$ correspond to the average population in the $\ket{m_s=0}$ state for the ensemble of spins.

Fig. \ref{stretch} shows a lifetime measurement on a static microdiamond following the protocol described in Sec.III. Here all four classes are resonant with the applied microwave frequency, which corresponds to the maximum degree of degeneracy between the NV centers, and therefore the stronger modification of the lifetime induced by the resonant dipolar coupling. The signal we obtain was fitted using a stretch exponential profile  and a simple exponential profile. We can see that the stretch exponential profile ($R^2=0.981$) is in better agreement with the data than the exponential fit ($R^2=0.942$). This is true in particular for the very short times (we expect the longer times to be dominated by the phonon-limited exponential lifetime).

Finally it should be noted that the stretch exponential profile arising from point-like depolarization sources is a relatively general result that has for example also been observed for the depolarization of NV centers induced by substitutional nitrogen (P1) defects in diamond \cite{hall_detection_2016_SI}

\subsection{Scanning the degeneracy conditions}
\begin{figure}
\includegraphics[scale=0.3]{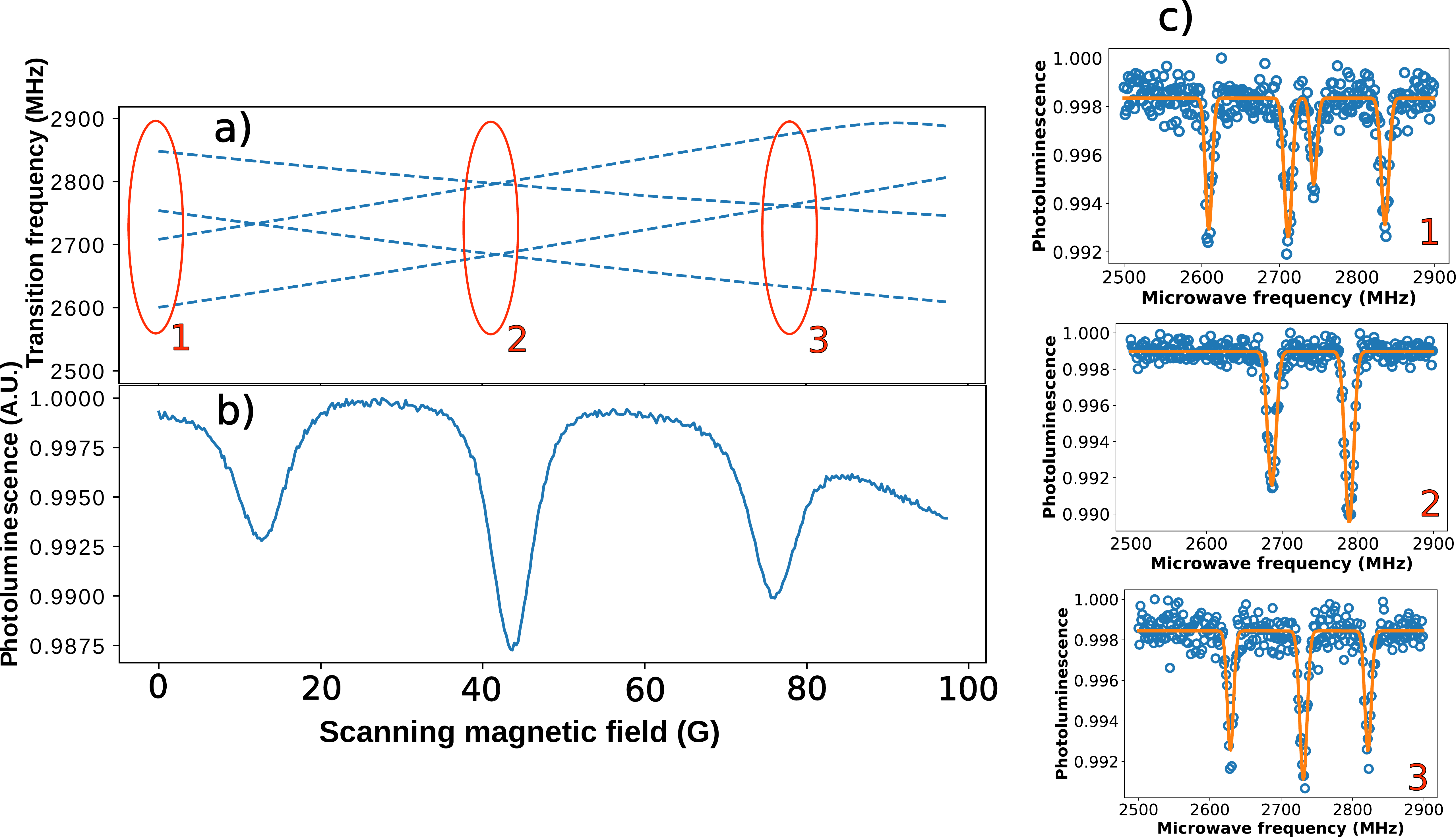}
\caption{Experimental results showing the PL versus magnetic field scans realized with an electromagnet. A static magnetic field offset is applied using a permanent magnet (setup similar to the one presented in Fig.\ref{Optics}) \textbf{a)} Frequency transitions on the $\ket{m_s=0} \to \ket{m_s=-1}$ electronic states of the four classes of NV centers as a function of the scanning magnetic field (data extrapolated from ODMR spectra) \textbf{b)} Change in photoluminescence of the NV$^-$ centers as a function of the scanning magnetic field. \textbf{c)} Example of 3 ODMR spectra for the 3 particular field values corresponding to the circled areas in subfigure a).}
	\label{scan_dege}
\end{figure}
The easiest way to probe the mechanism of dipolar-induced modification of the lifetime is to change the degeneracy conditions between the four classes of NV centers by tuning the magnetic field, as explained in Sec.I.
Because the NV spins can only exchange spin quanta when they are quasi-resonant, tuning the number of classes at degeneracy modifies the effective density of interacting NV centers, and therefore the depolarization effect. 

An example of this is given in Fig.2 of the main text with the varying lifetime depending on the degeneracy condition, but another way to probe this effect is shown in Fig.~\ref{scan_dege} : in this figure, we have observed the change in photoluminescence from a static microdiamond while changing the magnetic field in order to explore different degeneracy conditions. In order to do this, we need two sources of magnetic field : an electromagnet to scan the field and a permanent magnet to apply a magnetic field offset in an other direction (otherwise the magnetic field orientation with respect to the diamond axes would remain the same as the field is scanned).

In this particular case, we can see that as the magnetic field is scanned, it crosses three ``degeneracy planes" (as described in Sec.I) : first a plane of the the $\{ 110 \}$ family at B=13~G, with a single degeneracy condition, then a plane of the $\{ 100 \}$ family at B=44 G where there is a simultaneous degeneracy condition for two pairs of classes, and then another plane of the the $\{ 110 \}$ family at B=76~G. We notice that each time a degeneracy between at least two classes of NV takes place, a sharp decrease in photoluminescence is observed (see Fig.\ref{scan_dege}b)). This is a signature of the change in the lifetime of the ensemble of spins : indeed, the photoluminescence of NV ensembles is proportional to the average population in the $\ket{m_s=0}$ state, and the $\ket{m_s=0}$ population of the spins is the result of the competition between the polarization rate due to the green laser and the various depolarization mechanism. Increasing the depolarization rate of the spins will therefore decrease the overall luminosity.

\section{Experimental details}

\subsection{Experimental setup}

\begin{figure}[!ht]
  \centering \scalebox{0.3}{\includegraphics{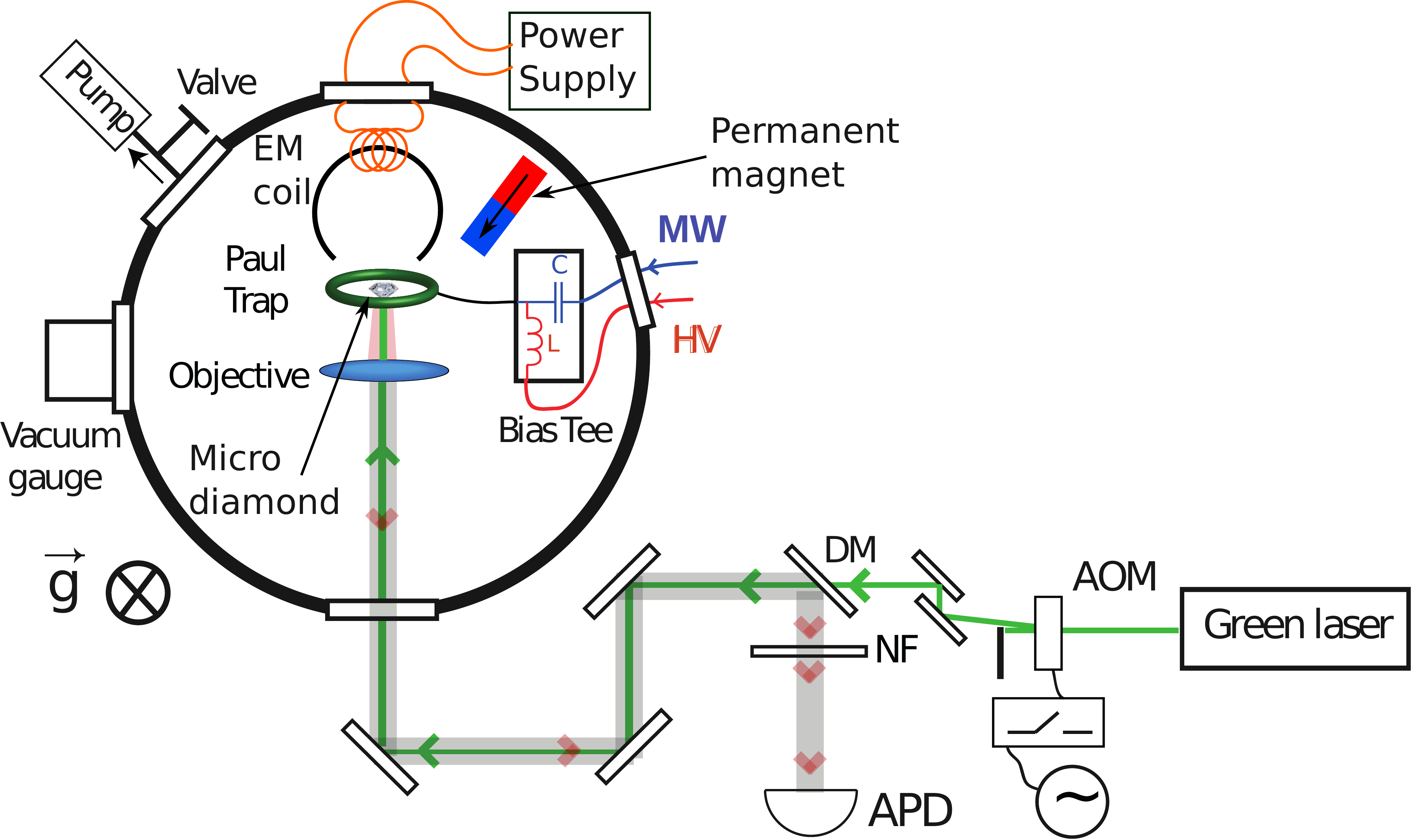}}
  \caption{Illustration of the experimental setup.}
	\label{Optics}
\end{figure}

The experimental setup illustrated in Fig.\ref{Optics} is similar to the one used in \citep{DelordPRL_SI} with the addition of a permanent magnet and an electromagnetic (EM) coil in order to perform magnetic field scans. The diamond sample is typically illuminated with 1mW of 532 nm laser light, focused by an objective with a numerical aperture of 0.5. An acousto-optic modulator (AOM) is used to switch on and off the 532nm laser and to finely tuned its power. The photo-luminescence (PL) is collected by the objective, separated form the excitation light using a dichroic mirror (DM) and a 532nm notch filter (NF), and detected using a multimode-fiber single-photon avalanche photo-detector (APD) (SPCM-AQRH-15 from Perkin Elmer). Typically, from the heavily doped samples that we use, we can detect PL photons at a rate of 1 MHz after attenuating the signal by a factor 100 with neutral density filters. 
The Paul trap is a pseudo-ring with a diameter of approximately 200 $\mu$m, as can be seen in \citep{DelordPhD_SI}. It acts both as trap through the high voltage (HV) and as a microwave (MW) antenna.

The magnetic field generated by the (homemade) EM coil is controlled by a programmable power supply (Rohde \& Schwarz NGE 103) performing current ramps.
While the levitating setup is located in a vacuum chamber, all the experiments presented in this article are performed at atmospheric pressure.

\subsection{$T_1$ measurement}
\begin{figure}[!ht]
  \centering \scalebox{1.2}{\includegraphics{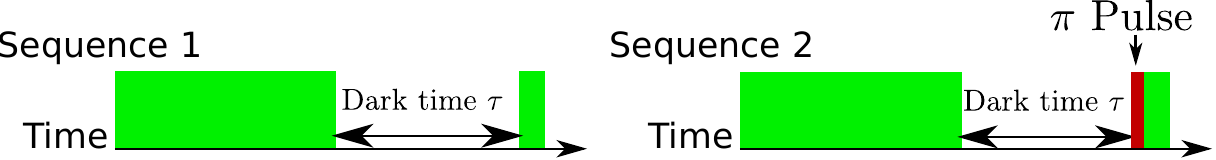}}
  \caption{$T_1$ measurement protocol. Green bars represent laser excitation, red bar represent resonant microwave $\pi$ pulse.}
	\label{T1_protocol}
\end{figure}

As shown in the Fig. 2-c) of the main text, the spin lifetime of the NV centers is modified in the presence of cross-relaxation with other classes of NV centers.
Here we present the protocol for removing the effects of charge state transfer in the dark, which mask the PL signal decay induced solely by spin depolarization. 
The protocol described in Fig. \ref{T1_protocol} consists in using two sequences. In the first sequence the spins are initially polarized in the $\ket{m_s=0}$ state through a 1 ms green laser excitation pulse and then left to evolve in the dark for a variable dark time $\tau$. The spin state is finally read out using a 10 $\mu$s laser pulse (shorter than the polarization time of the spins).

The second sequence uses the same parameters (polarization time, dark time and readout time) than the first sequence, but uses an extra resonant microwave $\pi$ pulse tuned to a transition of one of the four classes of NV$^-$ right before the readout pulse. The latter sequence brings population from the $\ket{m_s=0}$ state to the $\ket{m_s=\pm 1}$ state for one class of NV centers.

By measuring the difference between the two signals obtained in these two measurements, we can extract the evolution of the spin state population from a single NV class and, at the same time, remove unwanted contributions to the photoluminescence, such as charge state transfer in the dark (which give the same background contribution to the measurements).
In order to avoid low frequency noises such as laser drifts from the focal point or intensity fluctuations, we alternate both sequences while performing the measurement.

\subsection{Magnetic field calibration}

A neodymium permanent magnet and an electro-magnet are placed a few centimeters away from the diamond sample in order to apply a uniform and controllable magnetic field to the NV centers. 

To calibrate the magnetic field magnitude $B$, and its orientation $\theta$ with respect to the NV axis, we record Optically Detected Magnetic Resonance (ODMR) spectra and record the frequency of two transitions $|0 \rangle \rightarrow |-1\rangle$ and $|0 \rangle \rightarrow |+1\rangle$ from the same class to determine both the angle of the B field with respect to this class and the magnetic field amplitude. 

\subsection{Spin-mechanical detection}

High sensitivity of the spin-torque is achieved by using a speckle pattern produced by the rough surface of the micro-diamond under coherent illumination. When the particle is stably levitating, at the particle image plane, we then focus a small area of the speckle image onto an optical fibre and detect the photons transmitted through the fibre with the APD$_1$. The detected signal is then highly sensitive to the particle position and orientation.

For the spin-torque measurements presented in Fig. 4-a), the microwave detuning is scanned in 2~MHz steps with a duration of 10 ms per points. During those 10ms, the diamond orientation has enough time to reach its equilibrium position and the spin torque effect can be observed. The average count-rate is about 1 Mega-counts/s. 

\subsection{Angular signal drift for levitating particles}

Measurements on levitating diamonds have to be relatively short (few minutes at most) because of a slow drift on the particle orientation which changes the detection location on the specular reflection off the diamond surface. The most likely origin of this drift is the loss of charges of the diamond due to photoionization by the laser, which changes the trapping conditions over time.


%

\section{Principle of the mechanical detection}

\subsection{Origin of the magnetic torque}

The magnetic torque responsible for the motion of the diamond fundamentally comes from the anisotropy of the NV centers and from the transverse field $B_\perp$ responsible for mixing the eigenstates in the stationary state. We will start by considering the torque from a single NV center.
Without lack of generality, we will assume that the B field points in the $z$ direction and take the motion to be in the $x-z$ plane (in the lab frame), see Fig.~\ref{axes}.

\begin{figure}[ht]
\centering{ \scalebox{0.4} {\includegraphics{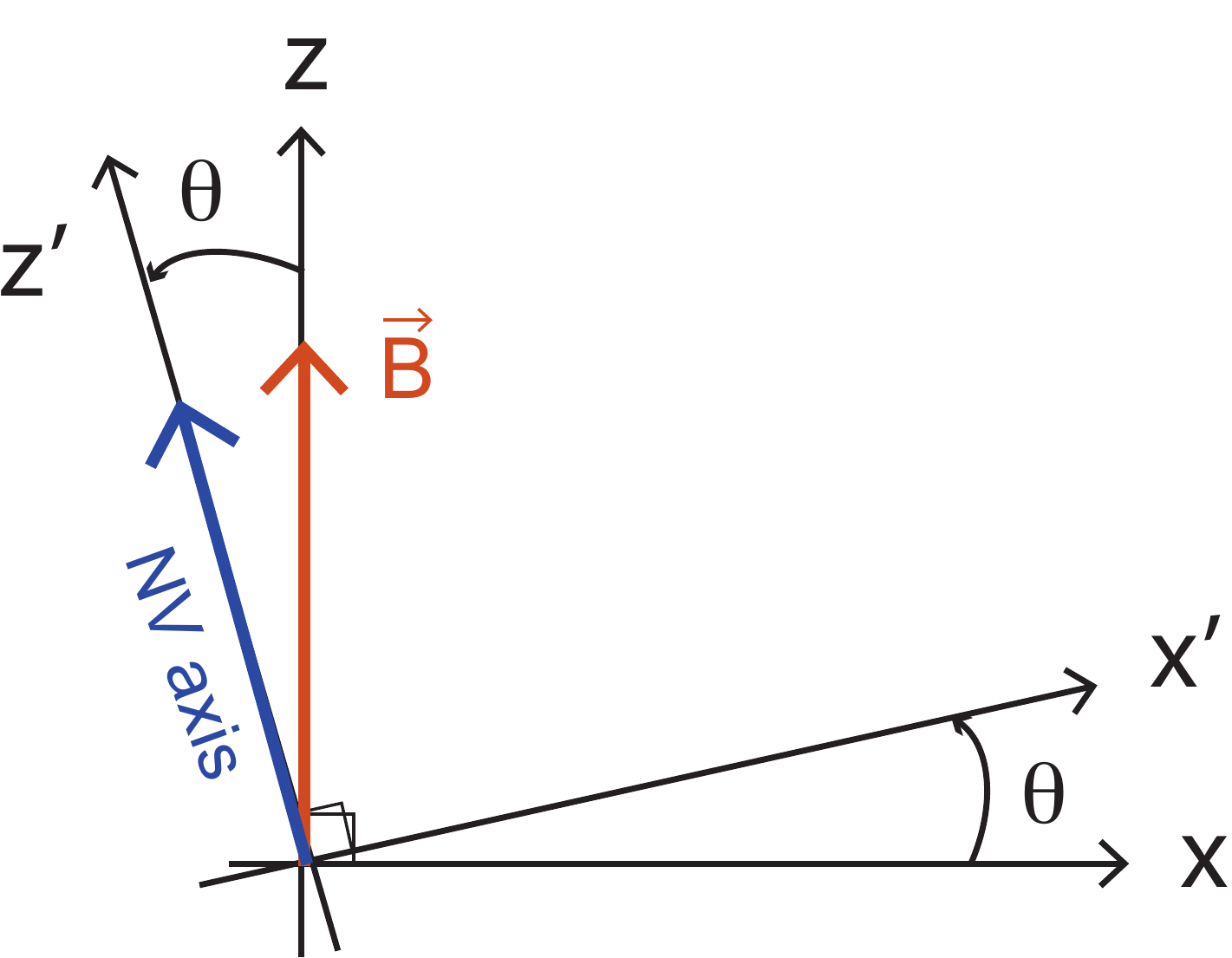}}}
  \caption{Notations used to define the axes in the body fixed and laboratory frames of reference $\mathcal{R'}$ and $\mathcal{R}$ respectively.}
  		\label{axes}
\end{figure}

In the body fixed frame, the magnetic part of the hamiltonian reads 
$\hat {H}_{B}= \hbar \gamma_e B  (\hat{S}_{x'} \sin\theta + \hat{S}_{z'} \cos\theta)$
where $\theta$ is the angle between the B field and NV center quantization axis $z'$.
We thus obtain the spin torque operator
\be
\hat \tau_s = -\frac{\partial \hat{H}}{\partial \theta} = \hbar \gamma_e B (-\cos\theta \hat{S}_{x'}+\sin\theta \hat{S}_{z'}).
\ee
The mean value of the torque operator in terms of the reduced density matrix elements $\rho_{ij}$ in the basis of the $\hat S_{z'}$ eigenstates $|-1\rangle_{z'},|0\rangle_{z'},|1\rangle_{z'}$ is 
\be
\langle \hat \tau_s \rangle =\mathrm{Tr}_\mathcal{B}(\hat \rho \hat \tau_s) = \hbar \gamma_e B   (\rho_{11}-\rho_{-1-1})\sin\theta - \hbar \frac{\gamma_e B}{\sqrt{2}} S \cos\theta,\label{torquemean}
\ee
where we introduced $S=\rho_{0,1}+\rho_{1,0}+\rho_{0,-1}+\rho_{-1,0}$.
The bath $\mathcal{B}$ over which the trace is performed consists of laser photons used to polarized the NV at a rate $\gamma_{\rm las}$, phonons or spin-fluctuators acting on the spin populations at a rate $\Gamma_1=1/T_1$ and 
$P_1$ centers or nuclear spins dephasing the electronic spin at a rate $1/T_2^*$.
In the limit $\gamma_e B \ll D$, and $\gamma_{\rm las}\gg \Gamma_1$ the laser efficiently polarizes the electronic spins in the ground state so that $\rho_{00}\gg {\rho_{11},\rho_{-1-1}}$.
The pure dephasing $T_2^*\approx 100$ns is much shorter than the sum of the population relaxation time $T_1/2 \le 1$ms and the laser induced repolarization time $1/\gamma_{\rm las} \le $ 100 $\mu$s. The equations of motion for the coherences thus read 

\bea
\frac{\partial \rho_{01}}{\partial t}&=&-\frac{1}{2T_2^*} \rho_{01}-i\frac{ \gamma_e B}{\sqrt{2}}\sin\theta-i\rho_{01}D+\mathcal{O}(\frac{(\gamma_e B)^2}{D})\\
\frac{\partial \rho_{0-1}}{\partial t}&=&-\frac{1}{2T_2^*}  \rho_{0-1}-i\frac{ \gamma_e B}{\sqrt{2}}\sin\theta-i\rho_{0-1}D+\mathcal{O}(\frac{(\gamma_e B)^2}{D}).
\eea
The characteristic motional dynamics is very slow compared to the zero-field and magnetic field rates $D$ and $\gamma B$. The latter are also much larger then the decoherence rate $1/T_2^*$ in our experiments, so we can adiabatically eliminate the coherences and find 
\bea
\rho_{01}=\rho_{10}\approx-\frac{\gamma_e B \sin\theta}{\sqrt{2} D} \quad {\rm and} \quad \rho_{0-1}=\rho_{-10}\approx-\frac{\gamma_e B \sin\theta}{\sqrt{2} D},\eea
since 
\bea
\rho_{11}-\rho_{-1-1}=\mathcal{O}((\frac{\gamma_e B}{D})^2).
\eea 
Re-injecting these expressions in the expression for the mean torque, we get 
\bea
\langle \hat \tau_s \rangle = \frac{\hbar(\gamma_e B)^2}{D}\sin2\theta
 +  \mathcal{O}(\frac{(\gamma_e B)^3}{D^2}).\label{torque_anal}
\eea

It is in fact the gradient of the energy $\partial / \partial \theta$ in the ground state at the angle $\theta$. 

Indeed, supposing that $\gamma B \ll D$, so that $\hat H_{B}$ can be treated as a perturbation to the spin-spin hamiltonian $D \hat S_{z'}^2$, 
the perturbed energy $\epsilon_0$ of $\ket{0}$ is 
\bea \epsilon_g=\sum_{m_s=\pm 1}\frac{ |\bra{0} H_B \ket{\pm 1}|^2}{-\epsilon^0_{\pm 1}}=-\hbar\frac{(\gamma_e B)^2 }{D} \sin^2\theta.
\eea

Taking $ -\partial \epsilon_g /\partial \theta$ then gives Eq. \ref{torque_anal}. It is the equation that is used in the core of the manuscript. 
In the approximate regime of the present study, the Hellmann-Feynman theorem (exact for pure states) that relates the angular derivative of the mean energy to the torque is correct in the above-described limits where dissipation is negligible. \\

\begin{figure}[ht]
\centering{ \scalebox{0.4} {\includegraphics{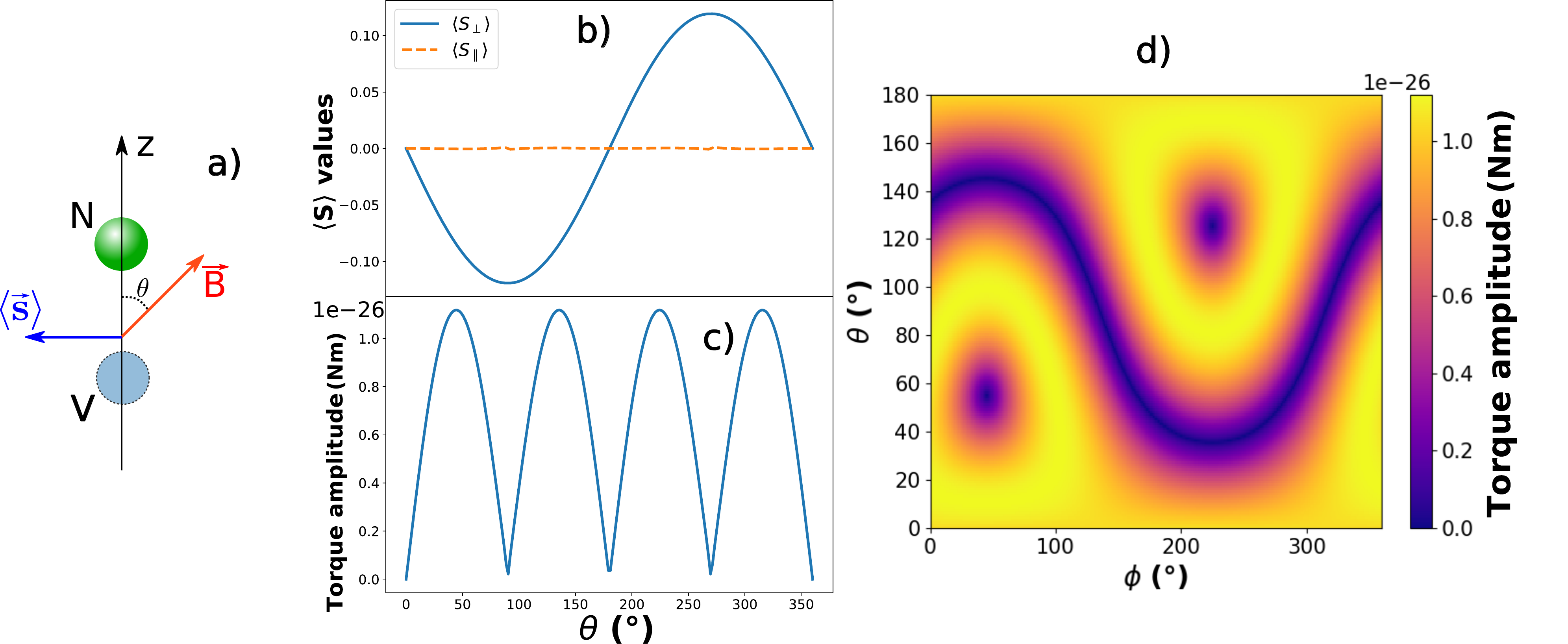}}}
  \caption{\textbf{a)} Sketch showing an NV center aligned to the \textbf{z'} direction in the presence of an external magnetic field \textbf{B} at an angle $\theta$ with respect the NV axis. The resulting spin vector $\langle \mathbf{ \hat S} \rangle$ of the NV center is shown by the blue arrow.
   \textbf{b)} Longitudinal ($S_\parallel$ in dashed line) and transverse ($S_\perp$ in plain line) components of the average value of the spin operator, in $\hbar$ unit, as a function of $\theta$ and for a magnetic field amplitude $|\mathbf{B}|=100$ G.   
   \textbf{c)} Amplitude of the magnetic torque acting on a single spin as a function of $\theta$ for $|\mathbf{B}|=100$ G.
   \textbf{d)} Amplitude of the same magnetic torque in the crystalline basis with $\theta$ and $\phi$ being the polar and azimuthal angle with respect to the [100] direction.}
  		\label{Torque1classe}
\end{figure}

Another way to estimate the torque is to numerically solve the master equation of the system as depicted in Fig \ref{Torque1classe}. We find that under green excitation and in the presence of an external magnetic field, the spins will acquire a magnetization $\gamma_e \langle\hat{\mathbf S}\rangle$ which, under the low magnetic fields ($<$ 200 G) we are working at, will be oriented at an angle of 90$^\circ$ from the NV axis : $\langle \hat S_{z'} \rangle \approx 0$ and $\langle \hat S_\perp \rangle \neq 0$. This magnetization vanishes when the magnetic field is aligned with the NV center since there is no longer a transverse field responsible for the mixing of the eigenstates. 

The magnetization of the NV center is therefore not aligned with the magnetic field, except when the field is also at a 90$^\circ$ angle from the NV axis, which means that the magnetic torque $\mathbf \Gamma = \gamma_e \langle\hat{\mathbf S}\rangle \times \mathbf B $ will be non-zero everywhere except when the field is aligned with the center, or in the plane normal to the direction of the center. We can describe each NV center as a paramagnetic defect with the anisotropic magnetic susceptibility $\chi = \begin{pmatrix}\chi_\perp & 0 & 0\\ 0 & \chi_\perp & 0 \\ 0 & 0 & 0  \end{pmatrix}$ in the $(\mathbf{x'},\mathbf{y'},\mathbf{z'})$ basis where $\mathbf{z'}$ is the orientation of the NV center.

The amplitude of the torque with respect to the magnetic field orientation at a B field amplitude of 100 G is represented in 1D in Fig \ref{Torque1classe}-c) where we can see a behavior very close to $|\sin(2\theta)|$, as found in Eq.\ref{torque_anal} through a perturbative approach. The same torque amplitude is represented in 2D in Fig \ref{Torque1classe} d). The two purple dots in the map correspond to the [111] direction when the magnetic field is aligned with the centers. The curvy line corresponds to the (111) plane. Importantly, the maximum torque value is $1\cdot 10^{-26}$ N.m for a single spin.

\subsection{Total spin torque and influence of cross-relaxation}


Fig 3 a) in the main text, represents the same map as Fig \ref{Torque1classe}-c) but including the four NV centers, one in each of the possible [111] orientations. We can see that the maximum torque actually decreased to $3\cdot 10^{-28}$ N.m even though we increased the number of NV centers by four. This is due to the directional averaging of the torque generated by the four centers. The torque per NV center is decreased by more than two orders of magnitude when taking the directional averaging into consideration.
Fig 3 b) in the main text shows the same map, this time taking into account the modification of the spin lifetime due to cross-relaxations. The detailed model is presented in the section \ref{Simu}.
There are two things to note here : 
\begin{enumerate}
\item The maximum torque has increased by an order of magnitude compared to the previous case. It reached up to $3\cdot 10^{-27}$ N.m for four spins, so about $10^{-27}$ N.m per spin. Qualitatively, this is because cross-relaxation will lower the torque contribution of specific classes (the ones that get depolarized), meaning that the end result is closer to the single spin case (there is less directional averaging).
\item The change in magnetic torque is resonant, and occurs only when different classes are brought to resonance. This can be seen by comparing Fig 3 b) of the main text to the $\{ 110 \}$ planes that were drawn in Fig. \ref{cristallo}. The change in the signal when scanning a magnetic field across a CR will be much sharper than the sinusoidal change in the spin-torque.
\end{enumerate} 

\subsection{Torque sensing with a levitating diamond}

The way we experimentally measure spin-torques applied on the levitating diamond is by measuring the induced diamond orientational displacement from equilibrium. 
We model the trap as a pure harmonic potential, both for the center of mass and for the librational degrees of freedom of the diamond with trapping frequencies $\omega_t \approx (2\pi) \cdot 1$ kHz. Considering a single librational degree of freedom, we can write the torque exerted by the trap as $\Gamma_t=-K(\theta-\theta_{eq})$, where $K=I \omega_t^2$ is the stiffness of the trap, $I$ being the moment of inertia of the diamond.

The application of an external torque $\Gamma_{\mathrm{ext}}$ to the diamond will therefore shift the angular equilibrium position  in such a way that : 
\begin{align}
-K(\theta-\theta_{\mathrm{eq}}) + \Gamma_{\mathrm{ext}} &= -K(\theta-\theta_{\mathrm{eq}}') 
\end{align}
so that
\begin{align}
\delta\theta = \theta_{\mathrm{eq}}'-\theta_{\mathrm{eq}} &= \frac{\Gamma_{\mathrm{ext}}}{K}=\frac{\Gamma_{\mathrm{ext}}}{I \omega_t^2}
\end{align}

In our case, $\Gamma_{\mathrm{ext}}$ is the magnetic torque exerted by the NV$^-$ spins on the diamond. We can write it $\Gamma_{\mathrm{ext}} = N_{NV} \langle \Gamma_{\mathrm{1 spin}} \rangle$  where $\langle \Gamma_{\mathrm{1 spin}} \rangle = \gamma_e \langle\hat{\mathbf S}\rangle \times \mathbf B \approx 10^{-27}$ Nm is the expected magnetic torque applied by one spin.

By using the inertia moment formula of a sphere : $I=\frac{2}{5}m r^2$, we can then rewrite the angular displacement as $$ \delta\theta = \frac{\langle\Gamma_{\mathrm{1 spin}}\rangle n(NV^-)}{\frac{2}{5}m_Cr^2\omega_T^2} \approx 10^{-3}~\rm rad$$ where $n(NV^-) \approx 5 \cdot 10^{-6}$ (5 ppm) is the number of NV centers per atoms in the crystal, $m_C \approx 2 \cdot 10^{-26}$ kg is the average weight of a carbon atom (we assume that the bulk of the diamond weight comes from carbon atoms), $r= 7.5$ $\mu$m is the typical radius of our diamonds and $\omega_T = 6.3\cdot 10^3$ rad/s is the typical value of the trap angular frequency.

It should be noted that the main uncertainty comes here from the diamond size, which can change the expected result by an order of magnitude.

\section{Cross-relaxation detection for another type of degeneracy}

\begin{figure}[!ht]
  \centering \scalebox{0.4}{\includegraphics{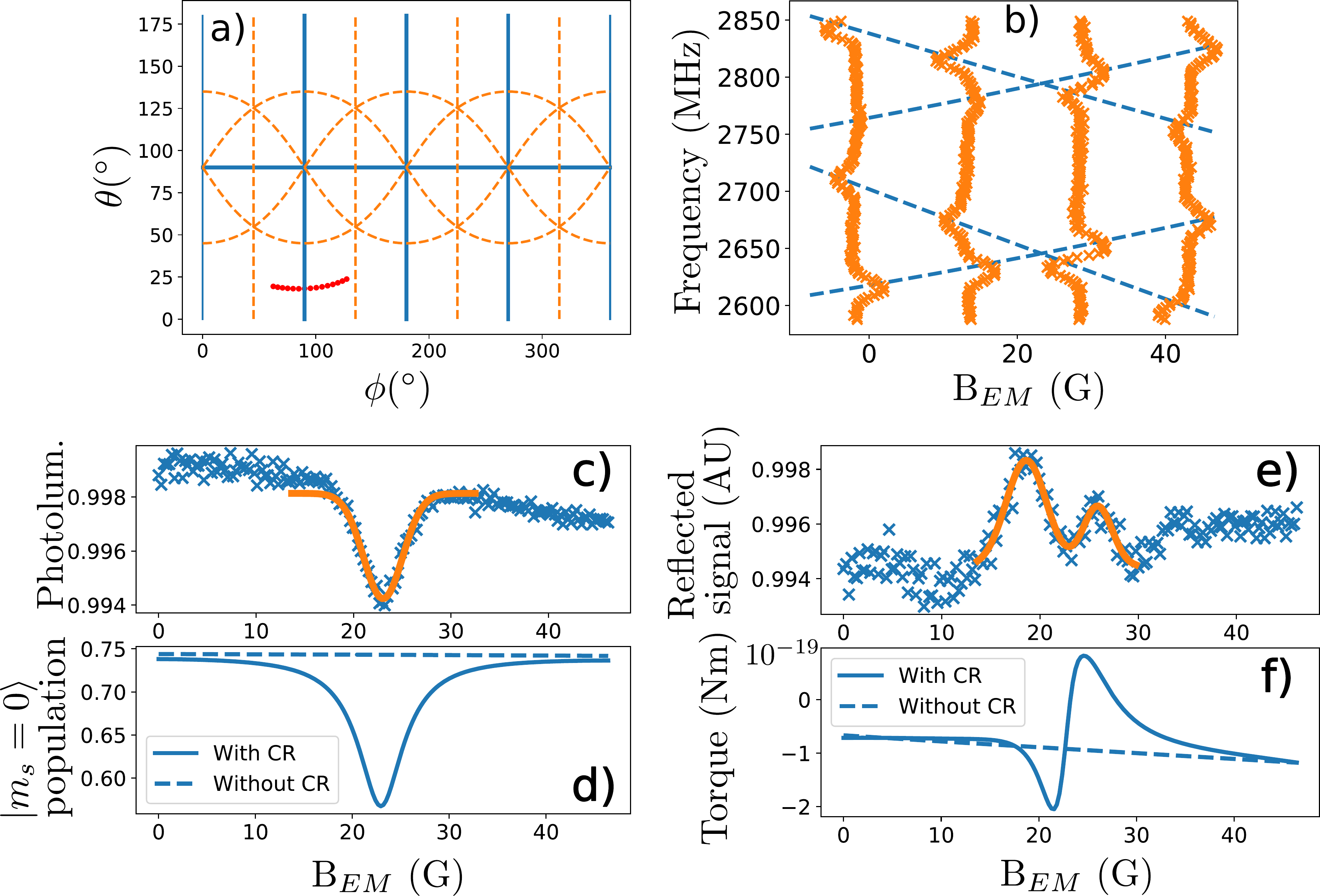}}
  \caption{Mechanical detection of a dipolar interaction when crossing a \{100\} plane. \textbf{a)} Path of the magnetic field angle (red dots) in the ($\theta , \phi$) basis in the measurements, with respect to the [100] direction. The \{100\}  family of planes are shown in plain blue lines. The \{110\}  family of planes is shown in orange dashed lines. \textbf{b)} ESR spectra measured at 4 different magnetic field values. Calculated evolution of the NV transition frequencies as a function of the electromagnet B field $B_{\rm EM}$, for the four $\ket{0} \to \ket{-1}$ transitions (dashed lines). \textbf{c)} Photoluminescence of the NV centers as a function of the scanning magnetic field (blue crosses) and gaussian fit (orange line). \textbf{d)} Simulated population in $\ket{m_s=0}$ for the stationary state with (plain) or without (dashed) taking into account the decrease of the T$_1$ induced by the cross-relaxations.
  \textbf{e)} Signal reflected off the diamond surface as a function of magnetic field amplitude $B_{\rm EM}$ (blue crosses). The orange line shows a double gaussian fit. \textbf{f)} Simulated torque applied by the spins on the diamond, with (plain) or without(dashed) taking into account the cross-relaxations.}
	\label{CR_22}
\end{figure}

Similarly to in Fig. 3 of the main text, we managed to mechanically detect other types of mechanically induced dipolar couplings.
Fig. \ref{CR_22}b) shows a calculation of the frequencies of the $\ket{0} \to \ket{-1}$ transitions for all four classes of NV centers and the mechanically detected ESR spectra measured using the reflected laser for various magnetic field values. Unlike the experiment reported in the main text, this time all four classes of NV are resonant with another class at B=23 G. This indicates that we are crossing a \{100\}  plane instead of a \{110\} plane, as can be seen in Fig \ref{CR_22}a).

Fig. \ref{CR_22}c) shows the recorded photoluminescence of the NV centers during the magnetic field scan. As expected, a drop in the photoluminescence rate is observed when the degeneracy occurs. The PL drop is slightly more pronounced in this case compared to the experiments presented in the main text because all classes are depolarized here, instead of only two. This is well predicted by the numerical estimates shown in Fig. \ref{CR_22}d).

Fig. \ref{CR_22}e) shows the signal of the laser reflected off the diamond surface, proportional to the angular displacement. Here, there is a clear difference between the angular response and the experiment shown in the main text. Instead of a single Gaussian drop centered on the CR resonance, two bumps are observed on both sides of the resonance. 
Fig. \ref{CR_22}f) shows the result of simulations, where we can see a dispersive profile with an almost zero torque at the resonance. 
The reason we do not observe a change of sign in the experiment (with two positive bumps instead of a positive and negative one) is likely to be because of the non-linearity of our detection : if the signal initially corresponds to a dark spot of the speckle, then a change in the motion of the diamond can only result in an increased signal. 

Let us give a physical interpretation of the dispersive angular profile for this degeneracy condition. 
Here, the magnetic torque generated by the four classes of NV is not modified exactly on resonance, since all four classes are depolarized identically. Close to resonance, all classes will not be identically depolarized however : looking at Fig. \ref{CR_22}b), we can see that the two classes which have a higher frequency are always slightly closer to each other than the two classes of lower frequency. This can be since by computing the slope in the evolution of the transitions frequencies, which are found to be smaller for the two upper classes. 
This effect results in more depolarization for these two classes, except when they are exactly on resonance. This interpretation explains the overall shape of the torque and thus of the angular response.
A quantitative analysis would require knowledge about the directions of the 4-NV directions with respect to the three main directions of the trapped diamond axes as well as a calibration of the sensitivity of the speckle detection method for the three corresponding angular modes. 

\section{Simulation details}
\label{Simu}
In this part we will discuss the method used to simulate the average torque as well as the population in the $\ket{m_s=0}$ state. Numerical solving of the master equation was performed using the Quantum Toolbox in Python (QuTiP) \citep{qutip1_SI} \citep{qutip2_SI}.

In order to describe the dynamics of our spin ensemble, we introduce a incoherent optical pumping through the jump operators $\mathcal{L}_+ = \Gamma_l \ket{0}\bra{+1} $ and $\mathcal{L}_- = \Gamma_l \ket{0}\bra{-1} $, where $\Gamma_l \approx (2\pi) 10$ kHz is the laser induced polarizing rate to the ground state.
We also introduce the $T_1$ jump operators $\mathcal{L}_i^j= \frac{1}{T_1}\ket{i}\bra{j}$ where $\ket{i,j}$=$\ket{0, \pm 1}$.

In order to describe the $T_1$ modification induced by the cross-relaxations, we use a phenomenological model where each class has its own T$_1^i$ ($i \in \{1,2,3,4\}$) that depends on the energy levels of the other classes with the formula :
\begin{equation*}
\frac{1}{T_1^i}=\frac{1}{T_1^0}+\sum_{j \neq i} \frac{1}{T_1^{dd}}e^{-\frac{(\nu_i-\nu_j)^2}{2(\sigma^{dd})^2}},
\end{equation*}
where $\nu_i$ and $\nu_j$ are the transition frequencies of the classes $i$ and $j$ (we are arbitrarily considering the $\ket{0} \to \ket{-1}$ transition here, since the resonance condition is the same for both transitions at the magnetic fields we are working at. This is not always true for magnetic fields greater than 592 G\citep{van_oort_cross-relaxation_1989_SI}). 
$\sigma^{dd} $ is the width of the CR features. We measured them to be similar to the inhomogeneous broadening given by the dipolar coupling to $P_1$ centers, {\it i.e.} $\approx$ 6 MHz.  

$T_1^0=1.03$ ms and $T_1^{dd}=0.38$ ms were chosen to match the $T_1$ measurements presented in Fig. 1 of the main text. We only focus on the $T_1$ without degeneracy and the one with a single degeneracy since our experiments will not have more than two resonant classes at once. Our model is probably not suited to deal with triple or quadruple resonances.

Finally, according to previous measurements performed in \citep{choi_depolarization_2017_SI}, only the $\ket{0}\bra{\pm1}$ and $\ket{\pm1}\bra{0}$ (corresponding to a single quantum exchange in the dipole-dipole interaction) operators are modified by the cross-relaxations. Using this model, we can numerically solve the master equation and get the density matrix in the stationary state $\rho_s$. With $\rho_s$ we can directly obtain the $\ket{m_s=0}$ population, corresponding to the experimentally measured photoluminescence.

With regards to the torque estimation, we use a semi-classical formula :
\begin{equation*}
\mathbf \Gamma = N_0 \gamma_e \langle\hat{\mathbf S}\rangle \times \mathbf B,
\end{equation*}
where $N_0 \approx 10^9$ is an estimate of the number of spins in our sample based on the average size and NV density of our diamonds, $\gamma_e$ is the gyromagnetic ratio of the electron and $\langle\hat{\mathbf S}\rangle =\mathrm{Tr}(\rho_s \mathbf{\hat S})$ is the averaged spin vector in the stationary sate, averaged again over the four possible orientations of NV.
This formula assumes that the spin dynamics is faster than the dynamics of the motion of the diamond, which is the case in our experiments.

In our plots in Fig. \ref{CR_22} and Fig. 3 of the main text, we only represent one spatial component (e.g. $\Gamma_x$) of the torque, because the three components behave similarly.

\end{widetext}


\begin{thebibliography}{47}%
\makeatletter
\providecommand \@ifxundefined [1]{%
 \@ifx{#1\undefined}
}%
\providecommand \@ifnum [1]{%
 \ifnum #1\expandafter \@firstoftwo
 \else \expandafter \@secondoftwo
 \fi
}%
\providecommand \@ifx [1]{%
 \ifx #1\expandafter \@firstoftwo
 \else \expandafter \@secondoftwo
 \fi
}%
\providecommand \natexlab [1]{#1}%
\providecommand \enquote  [1]{``#1''}%
\providecommand \bibnamefont  [1]{#1}%
\providecommand \bibfnamefont [1]{#1}%
\providecommand \citenamefont [1]{#1}%
\providecommand \href@noop [0]{\@secondoftwo}%
\providecommand \href [0]{\begingroup \@sanitize@url \@href}%
\providecommand \@href[1]{\@@startlink{#1}\@@href}%
\providecommand \@@href[1]{\endgroup#1\@@endlink}%
\providecommand \@sanitize@url [0]{\catcode `\\12\catcode `\$12\catcode
  `\&12\catcode `\#12\catcode `\^12\catcode `\_12\catcode `\%12\relax}%
\providecommand \@@startlink[1]{}%
\providecommand \@@endlink[0]{}%
\providecommand \url  [0]{\begingroup\@sanitize@url \@url }%
\providecommand \@url [1]{\endgroup\@href {#1}{\urlprefix }}%
\providecommand \urlprefix  [0]{URL }%
\providecommand \Eprint [0]{\href }%
\providecommand \doibase [0]{http://dx.doi.org/}%
\providecommand \selectlanguage [0]{\@gobble}%
\providecommand \bibinfo  [0]{\@secondoftwo}%
\providecommand \bibfield  [0]{\@secondoftwo}%
\providecommand \translation [1]{[#1]}%
\providecommand \BibitemOpen [0]{}%
\providecommand \bibitemStop [0]{}%
\providecommand \bibitemNoStop [0]{.\EOS\space}%
\providecommand \EOS [0]{\spacefactor3000\relax}%
\providecommand \BibitemShut  [1]{\csname bibitem#1\endcsname}%
\let\auto@bib@innerbib\@empty
\bibitem [{\citenamefont {Aspelmeyer}\ \emph {et~al.}(2014)\citenamefont
  {Aspelmeyer}, \citenamefont {Kippenberg},\ and\ \citenamefont
  {Marquardt}}]{Aspelmeyer}%
  \BibitemOpen
  \bibfield  {author} {\bibinfo {author} {\bibfnamefont {M.}~\bibnamefont
  {Aspelmeyer}}, \bibinfo {author} {\bibfnamefont {T.~J.}\ \bibnamefont
  {Kippenberg}}, \ and\ \bibinfo {author} {\bibfnamefont {F.}~\bibnamefont
  {Marquardt}},\ }\href@noop {} {\bibfield  {journal} {\bibinfo  {journal}
  {Rev. Mod. Phys.}\ }\textbf {\bibinfo {volume} {86}},\ \bibinfo {pages}
  {1391} (\bibinfo {year} {2014})}\BibitemShut {NoStop}%
\bibitem [{\citenamefont {Treutlein}\ \emph {et~al.}(2014)\citenamefont
  {Treutlein}, \citenamefont {Genes}, \citenamefont {Hammerer}, \citenamefont
  {Poggio},\ and\ \citenamefont {Rabl}}]{Treutlein2014}%
  \BibitemOpen
  \bibfield  {author} {\bibinfo {author} {\bibfnamefont {P.}~\bibnamefont
  {Treutlein}}, \bibinfo {author} {\bibfnamefont {C.}~\bibnamefont {Genes}},
  \bibinfo {author} {\bibfnamefont {K.}~\bibnamefont {Hammerer}}, \bibinfo
  {author} {\bibfnamefont {M.}~\bibnamefont {Poggio}}, \ and\ \bibinfo {author}
  {\bibfnamefont {P.}~\bibnamefont {Rabl}},\ }\enquote {\bibinfo {title}
  {Hybrid mechanical systems},}\ in\ \href {\doibase
  10.1007/978-3-642-55312-7_14} {\emph {\bibinfo {booktitle} {Cavity
  Optomechanics: Nano- and Micromechanical Resonators Interacting with
  Light}}},\ \bibinfo {editor} {edited by\ \bibinfo {editor} {\bibfnamefont
  {M.}~\bibnamefont {Aspelmeyer}}, \bibinfo {editor} {\bibfnamefont {T.~J.}\
  \bibnamefont {Kippenberg}}, \ and\ \bibinfo {editor} {\bibfnamefont
  {F.}~\bibnamefont {Marquardt}}}\ (\bibinfo  {publisher} {Springer Berlin
  Heidelberg},\ \bibinfo {address} {Berlin, Heidelberg},\ \bibinfo {year}
  {2014})\ pp.\ \bibinfo {pages} {327--351}\BibitemShut {NoStop}%
\bibitem [{\citenamefont {Rabl}\ \emph {et~al.}(2009)\citenamefont {Rabl},
  \citenamefont {Cappellaro}, \citenamefont {Dutt}, \citenamefont {Jiang},
  \citenamefont {Maze},\ and\ \citenamefont {Lukin}}]{Rabl}%
  \BibitemOpen
  \bibfield  {author} {\bibinfo {author} {\bibfnamefont {P.}~\bibnamefont
  {Rabl}}, \bibinfo {author} {\bibfnamefont {P.}~\bibnamefont {Cappellaro}},
  \bibinfo {author} {\bibfnamefont {M.~V.~G.}\ \bibnamefont {Dutt}}, \bibinfo
  {author} {\bibfnamefont {L.}~\bibnamefont {Jiang}}, \bibinfo {author}
  {\bibfnamefont {J.~R.}\ \bibnamefont {Maze}}, \ and\ \bibinfo {author}
  {\bibfnamefont {M.~D.}\ \bibnamefont {Lukin}},\ }\href@noop {} {\bibfield
  {journal} {\bibinfo  {journal} {Phys. Rev. B}\ }\textbf {\bibinfo {volume}
  {79}},\ \bibinfo {pages} {041302} (\bibinfo {year} {2009})}\BibitemShut
  {NoStop}%
\bibitem [{\citenamefont {Lee}\ \emph {et~al.}(2017)\citenamefont {Lee},
  \citenamefont {Lee}, \citenamefont {Cady}, \citenamefont {Ovartchaiyapong},\
  and\ \citenamefont {Jayich}}]{Lee_2017}%
  \BibitemOpen
  \bibfield  {author} {\bibinfo {author} {\bibfnamefont {D.}~\bibnamefont
  {Lee}}, \bibinfo {author} {\bibfnamefont {K.~W.}\ \bibnamefont {Lee}},
  \bibinfo {author} {\bibfnamefont {J.~V.}\ \bibnamefont {Cady}}, \bibinfo
  {author} {\bibfnamefont {P.}~\bibnamefont {Ovartchaiyapong}}, \ and\ \bibinfo
  {author} {\bibfnamefont {A.~C.~B.}\ \bibnamefont {Jayich}},\ }\href@noop {}
  {\bibfield  {journal} {\bibinfo  {journal} {Journal of Optics}\ }\textbf
  {\bibinfo {volume} {19}},\ \bibinfo {pages} {033001} (\bibinfo {year}
  {2017})}\BibitemShut {NoStop}%
\bibitem [{\citenamefont {Bose}\ \emph {et~al.}(2017)\citenamefont {Bose},
  \citenamefont {Mazumdar}, \citenamefont {Morley}, \citenamefont {Ulbricht},
  \citenamefont {Toro\ifmmode~\check{s}\else \v{s}\fi{}}, \citenamefont
  {Paternostro}, \citenamefont {Geraci}, \citenamefont {Barker}, \citenamefont
  {Kim},\ and\ \citenamefont {Milburn}}]{Bose}%
  \BibitemOpen
  \bibfield  {author} {\bibinfo {author} {\bibfnamefont {S.}~\bibnamefont
  {Bose}}, \bibinfo {author} {\bibfnamefont {A.}~\bibnamefont {Mazumdar}},
  \bibinfo {author} {\bibfnamefont {G.~W.}\ \bibnamefont {Morley}}, \bibinfo
  {author} {\bibfnamefont {H.}~\bibnamefont {Ulbricht}}, \bibinfo {author}
  {\bibfnamefont {M.}~\bibnamefont {Toro\ifmmode~\check{s}\else \v{s}\fi{}}},
  \bibinfo {author} {\bibfnamefont {M.}~\bibnamefont {Paternostro}}, \bibinfo
  {author} {\bibfnamefont {A.~A.}\ \bibnamefont {Geraci}}, \bibinfo {author}
  {\bibfnamefont {P.~F.}\ \bibnamefont {Barker}}, \bibinfo {author}
  {\bibfnamefont {M.~S.}\ \bibnamefont {Kim}}, \ and\ \bibinfo {author}
  {\bibfnamefont {G.}~\bibnamefont {Milburn}},\ }\href {\doibase
  10.1103/PhysRevLett.119.240401} {\bibfield  {journal} {\bibinfo  {journal}
  {Phys. Rev. Lett.}\ }\textbf {\bibinfo {volume} {119}},\ \bibinfo {pages}
  {240401} (\bibinfo {year} {2017})}\BibitemShut {NoStop}%
\bibitem [{\citenamefont {Wan}\ \emph {et~al.}(2016)\citenamefont {Wan},
  \citenamefont {Scala}, \citenamefont {Morley}, \citenamefont {Rahman},
  \citenamefont {Ulbricht}, \citenamefont {Bateman}, \citenamefont {Barker},
  \citenamefont {Bose},\ and\ \citenamefont {Kim}}]{Wan}%
  \BibitemOpen
  \bibfield  {author} {\bibinfo {author} {\bibfnamefont {C.}~\bibnamefont
  {Wan}}, \bibinfo {author} {\bibfnamefont {M.}~\bibnamefont {Scala}}, \bibinfo
  {author} {\bibfnamefont {G.~W.}\ \bibnamefont {Morley}}, \bibinfo {author}
  {\bibfnamefont {A.~A.}\ \bibnamefont {Rahman}}, \bibinfo {author}
  {\bibfnamefont {H.}~\bibnamefont {Ulbricht}}, \bibinfo {author}
  {\bibfnamefont {J.}~\bibnamefont {Bateman}}, \bibinfo {author} {\bibfnamefont
  {P.~F.}\ \bibnamefont {Barker}}, \bibinfo {author} {\bibfnamefont
  {S.}~\bibnamefont {Bose}}, \ and\ \bibinfo {author} {\bibfnamefont {M.~S.}\
  \bibnamefont {Kim}},\ }\href@noop {} {\bibfield  {journal} {\bibinfo
  {journal} {Phys. Rev. Lett.}\ }\textbf {\bibinfo {volume} {117}},\ \bibinfo
  {pages} {143003} (\bibinfo {year} {2016})}\BibitemShut {NoStop}%
\bibitem [{\citenamefont {Ma}\ \emph {et~al.}(2017)\citenamefont {Ma},
  \citenamefont {Hoang}, \citenamefont {Gong}, \citenamefont {Li},\ and\
  \citenamefont {Yin}}]{Ma}%
  \BibitemOpen
  \bibfield  {author} {\bibinfo {author} {\bibfnamefont {Y.}~\bibnamefont
  {Ma}}, \bibinfo {author} {\bibfnamefont {T.~M.}\ \bibnamefont {Hoang}},
  \bibinfo {author} {\bibfnamefont {M.}~\bibnamefont {Gong}}, \bibinfo {author}
  {\bibfnamefont {T.}~\bibnamefont {Li}}, \ and\ \bibinfo {author}
  {\bibfnamefont {Z.-q.}\ \bibnamefont {Yin}},\ }\href {\doibase
  10.1103/PhysRevA.96.023827} {\bibfield  {journal} {\bibinfo  {journal} {Phys.
  Rev. A}\ }\textbf {\bibinfo {volume} {96}},\ \bibinfo {pages} {023827}
  (\bibinfo {year} {2017})}\BibitemShut {NoStop}%
\bibitem [{\citenamefont {Kolkowitz}\ \emph {et~al.}(2012)\citenamefont
  {Kolkowitz}, \citenamefont {Bleszynski~Jayich}, \citenamefont
  {Unterreithmeier}, \citenamefont {Bennett}, \citenamefont {Rabl},
  \citenamefont {Harris},\ and\ \citenamefont {Lukin}}]{Kolkowitz}%
  \BibitemOpen
  \bibfield  {author} {\bibinfo {author} {\bibfnamefont {S.}~\bibnamefont
  {Kolkowitz}}, \bibinfo {author} {\bibfnamefont {A.~C.}\ \bibnamefont
  {Bleszynski~Jayich}}, \bibinfo {author} {\bibfnamefont {Q.~P.}\ \bibnamefont
  {Unterreithmeier}}, \bibinfo {author} {\bibfnamefont {S.~D.}\ \bibnamefont
  {Bennett}}, \bibinfo {author} {\bibfnamefont {P.}~\bibnamefont {Rabl}},
  \bibinfo {author} {\bibfnamefont {J.~G.~E.}\ \bibnamefont {Harris}}, \ and\
  \bibinfo {author} {\bibfnamefont {M.~D.}\ \bibnamefont {Lukin}},\ }\href
  {\doibase 10.1126/science.1216821} {\bibfield  {journal} {\bibinfo  {journal}
  {Science}\ }\textbf {\bibinfo {volume} {335}},\ \bibinfo {pages} {1603}
  (\bibinfo {year} {2012})}\BibitemShut {NoStop}%
\bibitem [{\citenamefont {Gieseler}\ \emph {et~al.}(2020)\citenamefont
  {Gieseler}, \citenamefont {Kabcenell}, \citenamefont {Rosenfeld},
  \citenamefont {Schaefer}, \citenamefont {Safira}, \citenamefont {Schuetz},
  \citenamefont {Gonzalez-Ballestero}, \citenamefont {Rusconi}, \citenamefont
  {Romero-Isart},\ and\ \citenamefont {Lukin}}]{Gieseler}%
  \BibitemOpen
  \bibfield  {author} {\bibinfo {author} {\bibfnamefont {J.}~\bibnamefont
  {Gieseler}}, \bibinfo {author} {\bibfnamefont {A.}~\bibnamefont {Kabcenell}},
  \bibinfo {author} {\bibfnamefont {E.}~\bibnamefont {Rosenfeld}}, \bibinfo
  {author} {\bibfnamefont {J.~D.}\ \bibnamefont {Schaefer}}, \bibinfo {author}
  {\bibfnamefont {A.}~\bibnamefont {Safira}}, \bibinfo {author} {\bibfnamefont
  {M.~J.~A.}\ \bibnamefont {Schuetz}}, \bibinfo {author} {\bibfnamefont
  {C.}~\bibnamefont {Gonzalez-Ballestero}}, \bibinfo {author} {\bibfnamefont
  {C.~C.}\ \bibnamefont {Rusconi}}, \bibinfo {author} {\bibfnamefont
  {O.}~\bibnamefont {Romero-Isart}}, \ and\ \bibinfo {author} {\bibfnamefont
  {M.~D.}\ \bibnamefont {Lukin}},\ }\href {\doibase
  10.1103/PhysRevLett.124.163604} {\bibfield  {journal} {\bibinfo  {journal}
  {Phys. Rev. Lett.}\ }\textbf {\bibinfo {volume} {124}},\ \bibinfo {pages}
  {163604} (\bibinfo {year} {2020})}\BibitemShut {NoStop}%
\bibitem [{\citenamefont {Delord}\ \emph {et~al.}(2020)\citenamefont {Delord},
  \citenamefont {Huillery}, \citenamefont {Nicolas},\ and\ \citenamefont
  {H{\'e}tet}}]{DelordNat}%
  \BibitemOpen
  \bibfield  {author} {\bibinfo {author} {\bibfnamefont {T.}~\bibnamefont
  {Delord}}, \bibinfo {author} {\bibfnamefont {P.}~\bibnamefont {Huillery}},
  \bibinfo {author} {\bibfnamefont {L.}~\bibnamefont {Nicolas}}, \ and\
  \bibinfo {author} {\bibfnamefont {G.}~\bibnamefont {H{\'e}tet}},\ }\href@noop
  {} {\bibfield  {journal} {\bibinfo  {journal} {Nature}\ }\textbf {\bibinfo
  {volume} {580}},\ \bibinfo {pages} {56} (\bibinfo {year} {2020})}\BibitemShut
  {NoStop}%
\bibitem [{\citenamefont {Arcizet}\ \emph {et~al.}(2011)\citenamefont
  {Arcizet}, \citenamefont {Jacques}, \citenamefont {Siria}, \citenamefont
  {Poncharal}, \citenamefont {Vincent},\ and\ \citenamefont
  {Seidelin}}]{Arcizet}%
  \BibitemOpen
  \bibfield  {author} {\bibinfo {author} {\bibfnamefont {O.}~\bibnamefont
  {Arcizet}}, \bibinfo {author} {\bibfnamefont {V.}~\bibnamefont {Jacques}},
  \bibinfo {author} {\bibfnamefont {A.}~\bibnamefont {Siria}}, \bibinfo
  {author} {\bibfnamefont {P.}~\bibnamefont {Poncharal}}, \bibinfo {author}
  {\bibfnamefont {P.}~\bibnamefont {Vincent}}, \ and\ \bibinfo {author}
  {\bibfnamefont {S.}~\bibnamefont {Seidelin}},\ }\href@noop {} {\bibfield
  {journal} {\bibinfo  {journal} {Nat Phys}\ }\textbf {\bibinfo {volume} {7}},\
  \bibinfo {pages} {879} (\bibinfo {year} {2011})}\BibitemShut {NoStop}%
\bibitem [{\citenamefont {Wei}\ \emph {et~al.}(2015)\citenamefont {Wei},
  \citenamefont {Burk}, \citenamefont {Wrachtrup},\ and\ \citenamefont
  {Liu}}]{Wei}%
  \BibitemOpen
  \bibfield  {author} {\bibinfo {author} {\bibfnamefont {B.-B.}\ \bibnamefont
  {Wei}}, \bibinfo {author} {\bibfnamefont {C.}~\bibnamefont {Burk}}, \bibinfo
  {author} {\bibfnamefont {J.}~\bibnamefont {Wrachtrup}}, \ and\ \bibinfo
  {author} {\bibfnamefont {R.-B.}\ \bibnamefont {Liu}},\ }\href@noop {}
  {\bibfield  {journal} {\bibinfo  {journal} {EPJ Quantum Technology}\ }\textbf
  {\bibinfo {volume} {2}},\ \bibinfo {pages} {18} (\bibinfo {year}
  {2015})}\BibitemShut {NoStop}%
\bibitem [{\citenamefont {Bachelard}\ \emph {et~al.}(2011)\citenamefont
  {Bachelard}, \citenamefont {Piovella},\ and\ \citenamefont
  {Courteille}}]{Bachelard}%
  \BibitemOpen
  \bibfield  {author} {\bibinfo {author} {\bibfnamefont {R.}~\bibnamefont
  {Bachelard}}, \bibinfo {author} {\bibfnamefont {N.}~\bibnamefont {Piovella}},
  \ and\ \bibinfo {author} {\bibfnamefont {P.~W.}\ \bibnamefont {Courteille}},\
  }\href {\doibase 10.1103/PhysRevA.84.013821} {\bibfield  {journal} {\bibinfo
  {journal} {Phys. Rev. A}\ }\textbf {\bibinfo {volume} {84}},\ \bibinfo
  {pages} {013821} (\bibinfo {year} {2011})}\BibitemShut {NoStop}%
\bibitem [{\citenamefont {Panat}\ and\ \citenamefont {Lawande}(2002)}]{PANAT}%
  \BibitemOpen
  \bibfield  {author} {\bibinfo {author} {\bibfnamefont {P.~V.}\ \bibnamefont
  {Panat}}\ and\ \bibinfo {author} {\bibfnamefont {S.~V.}\ \bibnamefont
  {Lawande}},\ }\href@noop {} {\bibfield  {journal} {\bibinfo  {journal}
  {International Journal of Modern Physics B}\ }\textbf {\bibinfo {volume}
  {16}},\ \bibinfo {pages} {3787} (\bibinfo {year} {2002})}\BibitemShut
  {NoStop}%
\bibitem [{\citenamefont {Prasanna~Venkatesh}\ \emph
  {et~al.}(2018)\citenamefont {Prasanna~Venkatesh}, \citenamefont {Juan},\ and\
  \citenamefont {Romero-Isart}}]{Venkatesh}%
  \BibitemOpen
  \bibfield  {author} {\bibinfo {author} {\bibfnamefont {B.}~\bibnamefont
  {Prasanna~Venkatesh}}, \bibinfo {author} {\bibfnamefont {M.~L.}\ \bibnamefont
  {Juan}}, \ and\ \bibinfo {author} {\bibfnamefont {O.}~\bibnamefont
  {Romero-Isart}},\ }\href {\doibase 10.1103/PhysRevLett.120.033602} {\bibfield
   {journal} {\bibinfo  {journal} {Phys. Rev. Lett.}\ }\textbf {\bibinfo
  {volume} {120}},\ \bibinfo {pages} {033602} (\bibinfo {year}
  {2018})}\BibitemShut {NoStop}%
\bibitem [{\citenamefont {Juan}\ \emph {et~al.}(2017)\citenamefont {Juan},
  \citenamefont {Bradac}, \citenamefont {Besga}, \citenamefont {Johnsson},
  \citenamefont {Brennen}, \citenamefont {Molina-Terriza},\ and\ \citenamefont
  {Volz}}]{Juan}%
  \BibitemOpen
  \bibfield  {author} {\bibinfo {author} {\bibfnamefont {M.~L.}\ \bibnamefont
  {Juan}}, \bibinfo {author} {\bibfnamefont {C.}~\bibnamefont {Bradac}},
  \bibinfo {author} {\bibfnamefont {B.}~\bibnamefont {Besga}}, \bibinfo
  {author} {\bibfnamefont {M.}~\bibnamefont {Johnsson}}, \bibinfo {author}
  {\bibfnamefont {G.}~\bibnamefont {Brennen}}, \bibinfo {author} {\bibfnamefont
  {G.}~\bibnamefont {Molina-Terriza}}, \ and\ \bibinfo {author} {\bibfnamefont
  {T.}~\bibnamefont {Volz}},\ }\href@noop {} {\bibfield  {journal} {\bibinfo
  {journal} {Nature Physics}\ }\textbf {\bibinfo {volume} {13}},\ \bibinfo
  {pages} {241} (\bibinfo {year} {2017})}\BibitemShut {NoStop}%
\bibitem [{\citenamefont {van Oort}\ and\ \citenamefont
  {Glasbeek}(1989)}]{van_oort_cross-relaxation_1989}%
  \BibitemOpen
  \bibfield  {author} {\bibinfo {author} {\bibfnamefont {E.}~\bibnamefont {van
  Oort}}\ and\ \bibinfo {author} {\bibfnamefont {M.}~\bibnamefont {Glasbeek}},\
  }\href {\doibase 10.1103/PhysRevB.40.6509} {\bibfield  {journal} {\bibinfo
  {journal} {Phys. Rev. B}\ }\textbf {\bibinfo {volume} {40}},\ \bibinfo
  {pages} {6509} (\bibinfo {year} {1989})},\ \bibinfo {note} {number:
  10}\BibitemShut {NoStop}%
\bibitem [{\citenamefont {Armstrong}\ \emph {et~al.}(2010)\citenamefont
  {Armstrong}, \citenamefont {Rogers}, \citenamefont {McMurtrie},\ and\
  \citenamefont {Manson}}]{armstrong_nvnv_2010}%
  \BibitemOpen
  \bibfield  {author} {\bibinfo {author} {\bibfnamefont {S.}~\bibnamefont
  {Armstrong}}, \bibinfo {author} {\bibfnamefont {L.~J.}\ \bibnamefont
  {Rogers}}, \bibinfo {author} {\bibfnamefont {R.~L.}\ \bibnamefont
  {McMurtrie}}, \ and\ \bibinfo {author} {\bibfnamefont {N.~B.}\ \bibnamefont
  {Manson}},\ }\href {\doibase 10.1016/j.phpro.2010.01.223} {\bibfield
  {journal} {\bibinfo  {journal} {Physics Procedia}\ }\textbf {\bibinfo
  {volume} {3}},\ \bibinfo {pages} {1569} (\bibinfo {year} {2010})},\ \bibinfo
  {note} {number: 4}\BibitemShut {NoStop}%
\bibitem [{\citenamefont {Alfasi}\ \emph {et~al.}(2019)\citenamefont {Alfasi},
  \citenamefont {Masis}, \citenamefont {Shtempluck},\ and\ \citenamefont
  {Buks}}]{Alfasi}%
  \BibitemOpen
  \bibfield  {author} {\bibinfo {author} {\bibfnamefont {N.}~\bibnamefont
  {Alfasi}}, \bibinfo {author} {\bibfnamefont {S.}~\bibnamefont {Masis}},
  \bibinfo {author} {\bibfnamefont {O.}~\bibnamefont {Shtempluck}}, \ and\
  \bibinfo {author} {\bibfnamefont {E.}~\bibnamefont {Buks}},\ }\href {\doibase
  10.1103/PhysRevB.99.214111} {\bibfield  {journal} {\bibinfo  {journal} {Phys.
  Rev. B}\ }\textbf {\bibinfo {volume} {99}},\ \bibinfo {pages} {214111}
  (\bibinfo {year} {2019})}\BibitemShut {NoStop}%
\bibitem [{\citenamefont {Epstein}\ \emph {et~al.}(2005)\citenamefont
  {Epstein}, \citenamefont {Mendoza}, \citenamefont {Kato},\ and\ \citenamefont
  {Awschalom}}]{Epstein}%
  \BibitemOpen
  \bibfield  {author} {\bibinfo {author} {\bibfnamefont {R.~J.}\ \bibnamefont
  {Epstein}}, \bibinfo {author} {\bibfnamefont {F.~M.}\ \bibnamefont
  {Mendoza}}, \bibinfo {author} {\bibfnamefont {Y.~K.}\ \bibnamefont {Kato}}, \
  and\ \bibinfo {author} {\bibfnamefont {D.~D.}\ \bibnamefont {Awschalom}},\
  }\href@noop {} {\bibfield  {journal} {\bibinfo  {journal} {Nature Physics}\
  }\textbf {\bibinfo {volume} {1}},\ \bibinfo {pages} {94} (\bibinfo {year}
  {2005})}\BibitemShut {NoStop}%
\bibitem [{\citenamefont {Hall}\ \emph {et~al.}(2016)\citenamefont {Hall},
  \citenamefont {Kehayias}, \citenamefont {Simpson}, \citenamefont {Jarmola},
  \citenamefont {Stacey}, \citenamefont {Budker},\ and\ \citenamefont
  {Hollenberg}}]{Hall}%
  \BibitemOpen
  \bibfield  {author} {\bibinfo {author} {\bibfnamefont {L.~T.}\ \bibnamefont
  {Hall}}, \bibinfo {author} {\bibfnamefont {P.}~\bibnamefont {Kehayias}},
  \bibinfo {author} {\bibfnamefont {D.~A.}\ \bibnamefont {Simpson}}, \bibinfo
  {author} {\bibfnamefont {A.}~\bibnamefont {Jarmola}}, \bibinfo {author}
  {\bibfnamefont {A.}~\bibnamefont {Stacey}}, \bibinfo {author} {\bibfnamefont
  {D.}~\bibnamefont {Budker}}, \ and\ \bibinfo {author} {\bibfnamefont
  {L.~C.~L.}\ \bibnamefont {Hollenberg}},\ }\href@noop {} {\bibfield  {journal}
  {\bibinfo  {journal} {Nature Communications}\ }\textbf {\bibinfo {volume}
  {7}},\ \bibinfo {pages} {10211} (\bibinfo {year} {2016})}\BibitemShut
  {NoStop}%
\bibitem [{\citenamefont {Wang}\ \emph {et~al.}(2013)\citenamefont {Wang},
  \citenamefont {Shin}, \citenamefont {Avalos}, \citenamefont {Seltzer},
  \citenamefont {Budker}, \citenamefont {Pines},\ and\ \citenamefont
  {Bajaj}}]{WangBajaj}%
  \BibitemOpen
  \bibfield  {author} {\bibinfo {author} {\bibfnamefont {H.-J.}\ \bibnamefont
  {Wang}}, \bibinfo {author} {\bibfnamefont {C.~S.}\ \bibnamefont {Shin}},
  \bibinfo {author} {\bibfnamefont {C.~E.}\ \bibnamefont {Avalos}}, \bibinfo
  {author} {\bibfnamefont {S.~J.}\ \bibnamefont {Seltzer}}, \bibinfo {author}
  {\bibfnamefont {D.}~\bibnamefont {Budker}}, \bibinfo {author} {\bibfnamefont
  {A.}~\bibnamefont {Pines}}, \ and\ \bibinfo {author} {\bibfnamefont {V.~S.}\
  \bibnamefont {Bajaj}},\ }\href {\doibase 10.1038/ncomms2930} {\bibfield
  {journal} {\bibinfo  {journal} {Nature Communications}\ }\textbf {\bibinfo
  {volume} {4}},\ \bibinfo {pages} {1940} (\bibinfo {year} {2013})}\BibitemShut
  {NoStop}%
\bibitem [{\citenamefont {Zangara}\ \emph {et~al.}(2019)\citenamefont
  {Zangara}, \citenamefont {Wood}, \citenamefont {Doherty},\ and\ \citenamefont
  {Meriles}}]{Zangara}%
  \BibitemOpen
  \bibfield  {author} {\bibinfo {author} {\bibfnamefont {P.~R.}\ \bibnamefont
  {Zangara}}, \bibinfo {author} {\bibfnamefont {A.}~\bibnamefont {Wood}},
  \bibinfo {author} {\bibfnamefont {M.~W.}\ \bibnamefont {Doherty}}, \ and\
  \bibinfo {author} {\bibfnamefont {C.~A.}\ \bibnamefont {Meriles}},\ }\href
  {\doibase 10.1103/PhysRevB.100.235410} {\bibfield  {journal} {\bibinfo
  {journal} {Phys. Rev. B}\ }\textbf {\bibinfo {volume} {100}},\ \bibinfo
  {pages} {235410} (\bibinfo {year} {2019})}\BibitemShut {NoStop}%
\bibitem [{\citenamefont {Abragam}(1989)}]{Abragam}%
  \BibitemOpen
  \bibfield  {author} {\bibinfo {author} {\bibfnamefont {A.}~\bibnamefont
  {Abragam}},\ }\href {https://cds.cern.ch/record/109195} {\emph {\bibinfo
  {title} {{The principles of nuclear magnetism; Reprint with corrections}}}},\
  International series of monographs on physics\ (\bibinfo  {publisher}
  {Clarendon Press},\ \bibinfo {address} {Oxford},\ \bibinfo {year}
  {1989})\BibitemShut {NoStop}%
\bibitem [{SM_()}]{SM_CR_meca}%
  \BibitemOpen
  \href@noop {} {\emph {\bibinfo {title} {See Supplemental Material at [...],
  which includes Refs. [26-34] and [17]}}}\BibitemShut {NoStop}%
\bibitem [{\citenamefont {Delord}\ \emph {et~al.}(2018)\citenamefont {Delord},
  \citenamefont {Huillery}, \citenamefont {Schwab}, \citenamefont {Nicolas},
  \citenamefont {Lecordier},\ and\ \citenamefont {H\'etet}}]{DelordPRL}%
  \BibitemOpen
  \bibfield  {author} {\bibinfo {author} {\bibfnamefont {T.}~\bibnamefont
  {Delord}}, \bibinfo {author} {\bibfnamefont {P.}~\bibnamefont {Huillery}},
  \bibinfo {author} {\bibfnamefont {L.}~\bibnamefont {Schwab}}, \bibinfo
  {author} {\bibfnamefont {L.}~\bibnamefont {Nicolas}}, \bibinfo {author}
  {\bibfnamefont {L.}~\bibnamefont {Lecordier}}, \ and\ \bibinfo {author}
  {\bibfnamefont {G.}~\bibnamefont {H\'etet}},\ }\href@noop {} {\bibfield
  {journal} {\bibinfo  {journal} {Phys. Rev. Lett.}\ }\textbf {\bibinfo
  {volume} {121}},\ \bibinfo {pages} {053602} (\bibinfo {year}
  {2018})}\BibitemShut {NoStop}%
\bibitem [{\citenamefont {Delord}(2019)}]{delordPhD}%
  \BibitemOpen
  \bibfield  {author} {\bibinfo {author} {\bibfnamefont {T.}~\bibnamefont
  {Delord}},\ }\emph {\bibinfo {title} {Spin-mechanics with micro-particles
  levitating in a Paul trap}},\ \href@noop {} {Ph.D. thesis},\ \bibinfo
  {school} {{\'E}cole Normale Sup{\'e}rieure} (\bibinfo {year}
  {2019})\BibitemShut {NoStop}%
\bibitem [{\citenamefont {Johansson}\ \emph {et~al.}(2012)\citenamefont
  {Johansson}, \citenamefont {Nation},\ and\ \citenamefont {Nori}}]{qutip1}%
  \BibitemOpen
  \bibfield  {author} {\bibinfo {author} {\bibfnamefont {J.~R.}\ \bibnamefont
  {Johansson}}, \bibinfo {author} {\bibfnamefont {P.~D.}\ \bibnamefont
  {Nation}}, \ and\ \bibinfo {author} {\bibfnamefont {F.}~\bibnamefont
  {Nori}},\ }\href@noop {} {\bibfield  {journal} {\bibinfo  {journal} {Computer
  Physics Communications}\ }\textbf {\bibinfo {volume} {183}},\ \bibinfo
  {pages} {1760} (\bibinfo {year} {2012})}\BibitemShut {NoStop}%
\bibitem [{\citenamefont {Johansson}(2013)}]{qutip2}%
  \BibitemOpen
  \bibfield  {author} {\bibinfo {author} {\bibfnamefont {J.}~\bibnamefont
  {Johansson}},\ }\href@noop {} {\bibfield  {journal} {\bibinfo  {journal}
  {Comp. Phys. Comm}\ }\textbf {\bibinfo {volume} {184}},\ \bibinfo {pages}
  {1234} (\bibinfo {year} {2013})}\BibitemShut {NoStop}%
\bibitem [{\citenamefont {Choi}\ \emph {et~al.}(2017)\citenamefont {Choi},
  \citenamefont {Choi}, \citenamefont {Kucsko}, \citenamefont {Maurer},
  \citenamefont {Shields}, \citenamefont {Sumiya}, \citenamefont {Onoda},
  \citenamefont {Isoya}, \citenamefont {Demler}, \citenamefont {Jelezko},
  \citenamefont {Yao},\ and\ \citenamefont {Lukin}}]{choi_depolarization_2017}%
  \BibitemOpen
  \bibfield  {author} {\bibinfo {author} {\bibfnamefont {J.}~\bibnamefont
  {Choi}}, \bibinfo {author} {\bibfnamefont {S.}~\bibnamefont {Choi}}, \bibinfo
  {author} {\bibfnamefont {G.}~\bibnamefont {Kucsko}}, \bibinfo {author}
  {\bibfnamefont {P.~C.}\ \bibnamefont {Maurer}}, \bibinfo {author}
  {\bibfnamefont {B.~J.}\ \bibnamefont {Shields}}, \bibinfo {author}
  {\bibfnamefont {H.}~\bibnamefont {Sumiya}}, \bibinfo {author} {\bibfnamefont
  {S.}~\bibnamefont {Onoda}}, \bibinfo {author} {\bibfnamefont
  {J.}~\bibnamefont {Isoya}}, \bibinfo {author} {\bibfnamefont
  {E.}~\bibnamefont {Demler}}, \bibinfo {author} {\bibfnamefont
  {F.}~\bibnamefont {Jelezko}}, \bibinfo {author} {\bibfnamefont {N.~Y.}\
  \bibnamefont {Yao}}, \ and\ \bibinfo {author} {\bibfnamefont {M.~D.}\
  \bibnamefont {Lukin}},\ }\href {\doibase 10.1103/PhysRevLett.118.093601}
  {\bibfield  {journal} {\bibinfo  {journal} {Phys. Rev. Lett.}\ }\textbf
  {\bibinfo {volume} {118}},\ \bibinfo {pages} {093601} (\bibinfo {year}
  {2017})},\ \bibinfo {note} {number: 9}\BibitemShut {NoStop}%
\bibitem [{\citenamefont {Jarmola}\ \emph {et~al.}(2012)\citenamefont
  {Jarmola}, \citenamefont {Acosta}, \citenamefont {Jensen}, \citenamefont
  {Chemerisov},\ and\ \citenamefont {Budker}}]{Jarmola}%
  \BibitemOpen
  \bibfield  {author} {\bibinfo {author} {\bibfnamefont {A.}~\bibnamefont
  {Jarmola}}, \bibinfo {author} {\bibfnamefont {V.~M.}\ \bibnamefont {Acosta}},
  \bibinfo {author} {\bibfnamefont {K.}~\bibnamefont {Jensen}}, \bibinfo
  {author} {\bibfnamefont {S.}~\bibnamefont {Chemerisov}}, \ and\ \bibinfo
  {author} {\bibfnamefont {D.}~\bibnamefont {Budker}},\ }\href@noop {}
  {\bibfield  {journal} {\bibinfo  {journal} {Phys. Rev. Lett.}\ }\textbf
  {\bibinfo {volume} {108}},\ \bibinfo {pages} {197601} (\bibinfo {year}
  {2012})}\BibitemShut {NoStop}%
\bibitem [{\citenamefont {Jarmola}\ \emph {et~al.}(2015)\citenamefont
  {Jarmola}, \citenamefont {Berzins}, \citenamefont {Smits}, \citenamefont
  {Smits}, \citenamefont {Prikulis}, \citenamefont {Gahbauer}, \citenamefont
  {Ferber}, \citenamefont {Erts}, \citenamefont {Auzinsh},\ and\ \citenamefont
  {Budker}}]{jarmola_longitudinal_2015}%
  \BibitemOpen
  \bibfield  {author} {\bibinfo {author} {\bibfnamefont {A.}~\bibnamefont
  {Jarmola}}, \bibinfo {author} {\bibfnamefont {A.}~\bibnamefont {Berzins}},
  \bibinfo {author} {\bibfnamefont {J.}~\bibnamefont {Smits}}, \bibinfo
  {author} {\bibfnamefont {K.}~\bibnamefont {Smits}}, \bibinfo {author}
  {\bibfnamefont {J.}~\bibnamefont {Prikulis}}, \bibinfo {author}
  {\bibfnamefont {F.}~\bibnamefont {Gahbauer}}, \bibinfo {author}
  {\bibfnamefont {R.}~\bibnamefont {Ferber}}, \bibinfo {author} {\bibfnamefont
  {D.}~\bibnamefont {Erts}}, \bibinfo {author} {\bibfnamefont {M.}~\bibnamefont
  {Auzinsh}}, \ and\ \bibinfo {author} {\bibfnamefont {D.}~\bibnamefont
  {Budker}},\ }\href {\doibase 10.1063/1.4937489} {\bibfield  {journal}
  {\bibinfo  {journal} {Appl. Phys. Lett.}\ }\textbf {\bibinfo {volume}
  {107}},\ \bibinfo {pages} {242403} (\bibinfo {year} {2015})},\ \bibinfo
  {note} {number: 24}\BibitemShut {NoStop}%
\bibitem [{\citenamefont {Akhmedzhanov}\ \emph {et~al.}(2017)\citenamefont
  {Akhmedzhanov}, \citenamefont {Gushchin}, \citenamefont {Nizov},
  \citenamefont {Nizov}, \citenamefont {Sobgayda}, \citenamefont {Zelensky},\
  and\ \citenamefont {Hemmer}}]{akhmedzhanov_microwave-free_2017}%
  \BibitemOpen
  \bibfield  {author} {\bibinfo {author} {\bibfnamefont {R.}~\bibnamefont
  {Akhmedzhanov}}, \bibinfo {author} {\bibfnamefont {L.}~\bibnamefont
  {Gushchin}}, \bibinfo {author} {\bibfnamefont {N.}~\bibnamefont {Nizov}},
  \bibinfo {author} {\bibfnamefont {V.}~\bibnamefont {Nizov}}, \bibinfo
  {author} {\bibfnamefont {D.}~\bibnamefont {Sobgayda}}, \bibinfo {author}
  {\bibfnamefont {I.}~\bibnamefont {Zelensky}}, \ and\ \bibinfo {author}
  {\bibfnamefont {P.}~\bibnamefont {Hemmer}},\ }\href {\doibase
  10.1103/PhysRevA.96.013806} {\bibfield  {journal} {\bibinfo  {journal} {Phys.
  Rev. A}\ }\textbf {\bibinfo {volume} {96}},\ \bibinfo {pages} {013806}
  (\bibinfo {year} {2017})},\ \bibinfo {note} {number: 1}\BibitemShut {NoStop}%
\bibitem [{\citenamefont {Akhmedzhanov}\ \emph {et~al.}(2019)\citenamefont
  {Akhmedzhanov}, \citenamefont {Gushchin}, \citenamefont {Nizov},
  \citenamefont {Nizov}, \citenamefont {Sobgayda}, \citenamefont {Zelensky},\
  and\ \citenamefont {Hemmer}}]{akhmedzhanov_magnetometry_2019}%
  \BibitemOpen
  \bibfield  {author} {\bibinfo {author} {\bibfnamefont {R.}~\bibnamefont
  {Akhmedzhanov}}, \bibinfo {author} {\bibfnamefont {L.}~\bibnamefont
  {Gushchin}}, \bibinfo {author} {\bibfnamefont {N.}~\bibnamefont {Nizov}},
  \bibinfo {author} {\bibfnamefont {V.}~\bibnamefont {Nizov}}, \bibinfo
  {author} {\bibfnamefont {D.}~\bibnamefont {Sobgayda}}, \bibinfo {author}
  {\bibfnamefont {I.}~\bibnamefont {Zelensky}}, \ and\ \bibinfo {author}
  {\bibfnamefont {P.}~\bibnamefont {Hemmer}},\ }\href {\doibase
  10.1103/PhysRevA.100.043844} {\bibfield  {journal} {\bibinfo  {journal}
  {Phys. Rev. A}\ }\textbf {\bibinfo {volume} {100}},\ \bibinfo {pages}
  {043844} (\bibinfo {year} {2019})},\ \bibinfo {note} {number: 4}\BibitemShut
  {NoStop}%
\bibitem [{\citenamefont {Holliday}\ \emph {et~al.}(1989)\citenamefont
  {Holliday}, \citenamefont {Manson}, \citenamefont {Glasbeek},\ and\
  \citenamefont {Oort}}]{holliday_optical_1989}%
  \BibitemOpen
  \bibfield  {author} {\bibinfo {author} {\bibfnamefont {K.}~\bibnamefont
  {Holliday}}, \bibinfo {author} {\bibfnamefont {N.~B.}\ \bibnamefont
  {Manson}}, \bibinfo {author} {\bibfnamefont {M.}~\bibnamefont {Glasbeek}}, \
  and\ \bibinfo {author} {\bibfnamefont {E.~v.}\ \bibnamefont {Oort}},\ }\href
  {\doibase 10.1088/0953-8984/1/39/021} {\bibfield  {journal} {\bibinfo
  {journal} {J. Phys.: Condens. Matter}\ }\textbf {\bibinfo {volume} {1}},\
  \bibinfo {pages} {7093} (\bibinfo {year} {1989})},\ \bibinfo {note} {number:
  39}\BibitemShut {NoStop}%
\bibitem [{\citenamefont {Mrózek}\ \emph {et~al.}(2015)\citenamefont
  {Mrózek}, \citenamefont {Rudnicki}, \citenamefont {Kehayias}, \citenamefont
  {Jarmola}, \citenamefont {Budker},\ and\ \citenamefont
  {Gawlik}}]{mrozek_longitudinal_2015}%
  \BibitemOpen
  \bibfield  {author} {\bibinfo {author} {\bibfnamefont {M.}~\bibnamefont
  {Mrózek}}, \bibinfo {author} {\bibfnamefont {D.}~\bibnamefont {Rudnicki}},
  \bibinfo {author} {\bibfnamefont {P.}~\bibnamefont {Kehayias}}, \bibinfo
  {author} {\bibfnamefont {A.}~\bibnamefont {Jarmola}}, \bibinfo {author}
  {\bibfnamefont {D.}~\bibnamefont {Budker}}, \ and\ \bibinfo {author}
  {\bibfnamefont {W.}~\bibnamefont {Gawlik}},\ }\href {\doibase
  10.1140/epjqt/s40507-015-0035-z} {\bibfield  {journal} {\bibinfo  {journal}
  {EPJ Quantum Technol.}\ }\textbf {\bibinfo {volume} {2}},\ \bibinfo {pages}
  {22} (\bibinfo {year} {2015})},\ \bibinfo {note} {number: 1}\BibitemShut
  {NoStop}%
\bibitem [{\citenamefont {Manson}\ \emph {et~al.}(2018)\citenamefont {Manson},
  \citenamefont {Hedges}, \citenamefont {Barson}, \citenamefont {Ahlefeldt},
  \citenamefont {Doherty}, \citenamefont {Abe}, \citenamefont {Ohshima},\ and\
  \citenamefont {Sellars}}]{manson_nv_2018}%
  \BibitemOpen
  \bibfield  {author} {\bibinfo {author} {\bibfnamefont {N.~B.}\ \bibnamefont
  {Manson}}, \bibinfo {author} {\bibfnamefont {M.}~\bibnamefont {Hedges}},
  \bibinfo {author} {\bibfnamefont {M.~S.~J.}\ \bibnamefont {Barson}}, \bibinfo
  {author} {\bibfnamefont {R.}~\bibnamefont {Ahlefeldt}}, \bibinfo {author}
  {\bibfnamefont {M.~W.}\ \bibnamefont {Doherty}}, \bibinfo {author}
  {\bibfnamefont {H.}~\bibnamefont {Abe}}, \bibinfo {author} {\bibfnamefont
  {T.}~\bibnamefont {Ohshima}}, \ and\ \bibinfo {author} {\bibfnamefont
  {M.~J.}\ \bibnamefont {Sellars}},\ }\href {\doibase 10.1088/1367-2630/aaec58}
  {\bibfield  {journal} {\bibinfo  {journal} {New J. Phys.}\ }\textbf {\bibinfo
  {volume} {20}},\ \bibinfo {pages} {113037} (\bibinfo {year}
  {2018})}\BibitemShut {NoStop}%
\bibitem [{\citenamefont {Tetienne}\ \emph {et~al.}(2013)\citenamefont
  {Tetienne}, \citenamefont {Hingant}, \citenamefont {Rondin}, \citenamefont
  {Cavaill\`es}, \citenamefont {Mayer}, \citenamefont {Dantelle}, \citenamefont
  {Gacoin}, \citenamefont {Wrachtrup}, \citenamefont {Roch},\ and\
  \citenamefont {Jacques}}]{Tetienne}%
  \BibitemOpen
  \bibfield  {author} {\bibinfo {author} {\bibfnamefont {J.-P.}\ \bibnamefont
  {Tetienne}}, \bibinfo {author} {\bibfnamefont {T.}~\bibnamefont {Hingant}},
  \bibinfo {author} {\bibfnamefont {L.}~\bibnamefont {Rondin}}, \bibinfo
  {author} {\bibfnamefont {A.}~\bibnamefont {Cavaill\`es}}, \bibinfo {author}
  {\bibfnamefont {L.}~\bibnamefont {Mayer}}, \bibinfo {author} {\bibfnamefont
  {G.}~\bibnamefont {Dantelle}}, \bibinfo {author} {\bibfnamefont
  {T.}~\bibnamefont {Gacoin}}, \bibinfo {author} {\bibfnamefont
  {J.}~\bibnamefont {Wrachtrup}}, \bibinfo {author} {\bibfnamefont {J.-F.}\
  \bibnamefont {Roch}}, \ and\ \bibinfo {author} {\bibfnamefont
  {V.}~\bibnamefont {Jacques}},\ }\href@noop {} {\bibfield  {journal} {\bibinfo
   {journal} {Phys. Rev. B}\ }\textbf {\bibinfo {volume} {87}},\ \bibinfo
  {pages} {235436} (\bibinfo {year} {2013})}\BibitemShut {NoStop}%
\bibitem [{\citenamefont {Dhomkar}\ \emph {et~al.}(2018)\citenamefont
  {Dhomkar}, \citenamefont {Jayakumar}, \citenamefont {Zangara},\ and\
  \citenamefont {Meriles}}]{Dhomkar}%
  \BibitemOpen
  \bibfield  {author} {\bibinfo {author} {\bibfnamefont {S.}~\bibnamefont
  {Dhomkar}}, \bibinfo {author} {\bibfnamefont {H.}~\bibnamefont {Jayakumar}},
  \bibinfo {author} {\bibfnamefont {P.~R.}\ \bibnamefont {Zangara}}, \ and\
  \bibinfo {author} {\bibfnamefont {C.~A.}\ \bibnamefont {Meriles}},\
  }\href@noop {} {\bibfield  {journal} {\bibinfo  {journal} {Nano Letters}\
  }\textbf {\bibinfo {volume} {18}},\ \bibinfo {pages} {4046} (\bibinfo {year}
  {2018})}\BibitemShut {NoStop}%
\bibitem [{\citenamefont {Giri}\ \emph {et~al.}(2019)\citenamefont {Giri},
  \citenamefont {Dorigoni}, \citenamefont {Tambalo}, \citenamefont {Gorrini},\
  and\ \citenamefont {Bifone}}]{giri_selective_2019}%
  \BibitemOpen
  \bibfield  {author} {\bibinfo {author} {\bibfnamefont {R.}~\bibnamefont
  {Giri}}, \bibinfo {author} {\bibfnamefont {C.}~\bibnamefont {Dorigoni}},
  \bibinfo {author} {\bibfnamefont {S.}~\bibnamefont {Tambalo}}, \bibinfo
  {author} {\bibfnamefont {F.}~\bibnamefont {Gorrini}}, \ and\ \bibinfo
  {author} {\bibfnamefont {A.}~\bibnamefont {Bifone}},\ }\href {\doibase
  10.1103/PhysRevB.99.155426} {\bibfield  {journal} {\bibinfo  {journal} {Phys.
  Rev. B}\ }\textbf {\bibinfo {volume} {99}},\ \bibinfo {pages} {155426}
  (\bibinfo {year} {2019})},\ \bibinfo {note} {number: 15}\BibitemShut
  {NoStop}%
\bibitem [{\citenamefont {Giri}\ \emph {et~al.}(2018)\citenamefont {Giri},
  \citenamefont {Gorrini}, \citenamefont {Dorigoni}, \citenamefont {Avalos},
  \citenamefont {Cazzanelli}, \citenamefont {Tambalo},\ and\ \citenamefont
  {Bifone}}]{giri_coupled_2018}%
  \BibitemOpen
  \bibfield  {author} {\bibinfo {author} {\bibfnamefont {R.}~\bibnamefont
  {Giri}}, \bibinfo {author} {\bibfnamefont {F.}~\bibnamefont {Gorrini}},
  \bibinfo {author} {\bibfnamefont {C.}~\bibnamefont {Dorigoni}}, \bibinfo
  {author} {\bibfnamefont {C.~E.}\ \bibnamefont {Avalos}}, \bibinfo {author}
  {\bibfnamefont {M.}~\bibnamefont {Cazzanelli}}, \bibinfo {author}
  {\bibfnamefont {S.}~\bibnamefont {Tambalo}}, \ and\ \bibinfo {author}
  {\bibfnamefont {A.}~\bibnamefont {Bifone}},\ }\href {\doibase
  10.1103/PhysRevB.98.045401} {\bibfield  {journal} {\bibinfo  {journal} {Phys.
  Rev. B}\ }\textbf {\bibinfo {volume} {98}},\ \bibinfo {pages} {045401}
  (\bibinfo {year} {2018})},\ \bibinfo {note} {number: 4}\BibitemShut {NoStop}%
\bibitem [{\citenamefont {Rugar}\ \emph {et~al.}(2004)\citenamefont {Rugar},
  \citenamefont {Budakian}, \citenamefont {Mamin},\ and\ \citenamefont
  {Chui}}]{Rugar}%
  \BibitemOpen
  \bibfield  {author} {\bibinfo {author} {\bibfnamefont {D.}~\bibnamefont
  {Rugar}}, \bibinfo {author} {\bibfnamefont {R.}~\bibnamefont {Budakian}},
  \bibinfo {author} {\bibfnamefont {H.~J.}\ \bibnamefont {Mamin}}, \ and\
  \bibinfo {author} {\bibfnamefont {B.~W.}\ \bibnamefont {Chui}},\ }\href
  {https://doi.org/10.1038/nature02658} {\bibfield  {journal} {\bibinfo
  {journal} {Nature}\ }\textbf {\bibinfo {volume} {430}},\ \bibinfo {pages}
  {329 EP } (\bibinfo {year} {2004})}\BibitemShut {NoStop}%
\bibitem [{\citenamefont {Delord}\ \emph {et~al.}(2017)\citenamefont {Delord},
  \citenamefont {Nicolas}, \citenamefont {Bodini},\ and\ \citenamefont
  {H{\'e}tet}}]{vacuumESR}%
  \BibitemOpen
  \bibfield  {author} {\bibinfo {author} {\bibfnamefont {T.}~\bibnamefont
  {Delord}}, \bibinfo {author} {\bibfnamefont {L.}~\bibnamefont {Nicolas}},
  \bibinfo {author} {\bibfnamefont {M.}~\bibnamefont {Bodini}}, \ and\ \bibinfo
  {author} {\bibfnamefont {G.}~\bibnamefont {H{\'e}tet}},\ }\href@noop {}
  {\bibfield  {journal} {\bibinfo  {journal} {Applied Physics Letters}\
  }\textbf {\bibinfo {volume} {111}},\ \bibinfo {pages} {013101} (\bibinfo
  {year} {2017})}\BibitemShut {NoStop}%
\bibitem [{\citenamefont {{Braginski{\v{i}}}}\ and\ \citenamefont
  {{Manukin}}(1967)}]{Braginski}%
  \BibitemOpen
  \bibfield  {author} {\bibinfo {author} {\bibfnamefont {V.~B.}\ \bibnamefont
  {{Braginski{\v{i}}}}}\ and\ \bibinfo {author} {\bibfnamefont {A.~B.}\
  \bibnamefont {{Manukin}}},\ }\href@noop {} {\bibfield  {journal} {\bibinfo
  {journal} {Soviet Journal of Experimental and Theoretical Physics}\ }\textbf
  {\bibinfo {volume} {25}},\ \bibinfo {pages} {653} (\bibinfo {year}
  {1967})}\BibitemShut {NoStop}%
\bibitem [{\citenamefont {Kim}\ \emph {et~al.}(2016)\citenamefont {Kim},
  \citenamefont {Hauer}, \citenamefont {Doolin}, \citenamefont {Souris},\ and\
  \citenamefont {Davis}}]{Kim}%
  \BibitemOpen
  \bibfield  {author} {\bibinfo {author} {\bibfnamefont {P.~H.}\ \bibnamefont
  {Kim}}, \bibinfo {author} {\bibfnamefont {B.~D.}\ \bibnamefont {Hauer}},
  \bibinfo {author} {\bibfnamefont {C.}~\bibnamefont {Doolin}}, \bibinfo
  {author} {\bibfnamefont {F.}~\bibnamefont {Souris}}, \ and\ \bibinfo {author}
  {\bibfnamefont {J.~P.}\ \bibnamefont {Davis}},\ }\href@noop {} {\bibfield
  {journal} {\bibinfo  {journal} {Nature Communications}\ }\textbf {\bibinfo
  {volume} {7}},\ \bibinfo {pages} {13165 EP } (\bibinfo {year}
  {2016})}\BibitemShut {NoStop}%
\bibitem [{\citenamefont {Ahn}\ \emph {et~al.}(2020)\citenamefont {Ahn},
  \citenamefont {Xu}, \citenamefont {Bang}, \citenamefont {Ju}, \citenamefont
  {Gao},\ and\ \citenamefont {Li}}]{Jonghoon}%
  \BibitemOpen
  \bibfield  {author} {\bibinfo {author} {\bibfnamefont {J.}~\bibnamefont
  {Ahn}}, \bibinfo {author} {\bibfnamefont {Z.}~\bibnamefont {Xu}}, \bibinfo
  {author} {\bibfnamefont {J.}~\bibnamefont {Bang}}, \bibinfo {author}
  {\bibfnamefont {P.}~\bibnamefont {Ju}}, \bibinfo {author} {\bibfnamefont
  {X.}~\bibnamefont {Gao}}, \ and\ \bibinfo {author} {\bibfnamefont
  {T.}~\bibnamefont {Li}},\ }\href@noop {} {\bibfield  {journal} {\bibinfo
  {journal} {Nature Nanotechnology}\ }\textbf {\bibinfo {volume} {15}},\
  \bibinfo {pages} {89} (\bibinfo {year} {2020})}\BibitemShut {NoStop}%
\bibitem [{\citenamefont {Carlin}(2012)}]{carlin}%
  \BibitemOpen
  \bibfield  {author} {\bibinfo {author} {\bibfnamefont {R.~L.}\ \bibnamefont
  {Carlin}},\ }\href@noop {} {\emph {\bibinfo {title} {Magnetochemistry}}}\
  (\bibinfo  {publisher} {Springer Science \& Business Media},\ \bibinfo {year}
  {2012})\BibitemShut {NoStop}%
\end{thebibliography}

\begin{thebibliography}{11}%
\makeatletter
\providecommand \@ifxundefined [1]{%
 \@ifx{#1\undefined}
}%
\providecommand \@ifnum [1]{%
 \ifnum #1\expandafter \@firstoftwo
 \else \expandafter \@secondoftwo
 \fi
}%
\providecommand \@ifx [1]{%
 \ifx #1\expandafter \@firstoftwo
 \else \expandafter \@secondoftwo
 \fi
}%
\providecommand \natexlab [1]{#1}%
\providecommand \enquote  [1]{``#1''}%
\providecommand \bibnamefont  [1]{#1}%
\providecommand \bibfnamefont [1]{#1}%
\providecommand \citenamefont [1]{#1}%
\providecommand \href@noop [0]{\@secondoftwo}%
\providecommand \href [0]{\begingroup \@sanitize@url \@href}%
\providecommand \@href[1]{\@@startlink{#1}\@@href}%
\providecommand \@@href[1]{\endgroup#1\@@endlink}%
\providecommand \@sanitize@url [0]{\catcode `\\12\catcode `\$12\catcode
  `\&12\catcode `\#12\catcode `\^12\catcode `\_12\catcode `\%12\relax}%
\providecommand \@@startlink[1]{}%
\providecommand \@@endlink[0]{}%
\providecommand \url  [0]{\begingroup\@sanitize@url \@url }%
\providecommand \@url [1]{\endgroup\@href {#1}{\urlprefix }}%
\providecommand \urlprefix  [0]{URL }%
\providecommand \Eprint [0]{\href }%
\providecommand \doibase [0]{http://dx.doi.org/}%
\providecommand \selectlanguage [0]{\@gobble}%
\providecommand \bibinfo  [0]{\@secondoftwo}%
\providecommand \bibfield  [0]{\@secondoftwo}%
\providecommand \translation [1]{[#1]}%
\providecommand \BibitemOpen [0]{}%
\providecommand \bibitemStop [0]{}%
\providecommand \bibitemNoStop [0]{.\EOS\space}%
\providecommand \EOS [0]{\spacefactor3000\relax}%
\providecommand \BibitemShut  [1]{\csname bibitem#1\endcsname}%
\let\auto@bib@innerbib\@empty

\bibitem [{\citenamefont {Jarmola}\ \emph {et~al.}(2012)\citenamefont
  {Jarmola}, \citenamefont {Acosta}, \citenamefont {Jensen}, \citenamefont
  {Chemerisov},\ and\ \citenamefont {Budker}}]{Jarmola_SI}%
  \BibitemOpen
  \bibfield  {author} {\bibinfo {author} {\bibfnamefont {A.}~\bibnamefont
  {Jarmola}}, \bibinfo {author} {\bibfnamefont {V.~M.}\ \bibnamefont {Acosta}},
  \bibinfo {author} {\bibfnamefont {K.}~\bibnamefont {Jensen}}, \bibinfo
  {author} {\bibfnamefont {S.}~\bibnamefont {Chemerisov}}, \ and\ \bibinfo
  {author} {\bibfnamefont {D.}~\bibnamefont {Budker}},\ }\href@noop {}
  {\bibfield  {journal} {\bibinfo  {journal} {Phys. Rev. Lett.}\ }\textbf
  {\bibinfo {volume} {108}},\ \bibinfo {pages} {197601} (\bibinfo {year}
  {2012})}\BibitemShut {NoStop}%
\bibitem [{\citenamefont {Mrózek}\ \emph {et~al.}(2015)\citenamefont
  {Mrózek}, \citenamefont {Rudnicki}, \citenamefont {Kehayias}, \citenamefont
  {Jarmola}, \citenamefont {Budker},\ and\ \citenamefont
  {Gawlik}}]{mrozek_longitudinal_2015_SI}%
  \BibitemOpen
  \bibfield  {author} {\bibinfo {author} {\bibfnamefont {M.}~\bibnamefont
  {Mrózek}}, \bibinfo {author} {\bibfnamefont {D.}~\bibnamefont {Rudnicki}},
  \bibinfo {author} {\bibfnamefont {P.}~\bibnamefont {Kehayias}}, \bibinfo
  {author} {\bibfnamefont {A.}~\bibnamefont {Jarmola}}, \bibinfo {author}
  {\bibfnamefont {D.}~\bibnamefont {Budker}}, \ and\ \bibinfo {author}
  {\bibfnamefont {W.}~\bibnamefont {Gawlik}},\ }\href {\doibase
  10.1140/epjqt/s40507-015-0035-z} {\bibfield  {journal} {\bibinfo  {journal}
  {EPJ Quantum Technol.}\ }\textbf {\bibinfo {volume} {2}},\ \bibinfo {pages}
  {22} (\bibinfo {year} {2015})},\ \bibinfo {note} {number: 1}\BibitemShut
  {NoStop}%
\bibitem [{\citenamefont {Choi}\ \emph {et~al.}(2017)\citenamefont {Choi},
  \citenamefont {Choi}, \citenamefont {Kucsko}, \citenamefont {Maurer},
  \citenamefont {Shields}, \citenamefont {Sumiya}, \citenamefont {Onoda},
  \citenamefont {Isoya}, \citenamefont {Demler}, \citenamefont {Jelezko},
  \citenamefont {Yao},\ and\ \citenamefont {Lukin}}]{choi_depolarization_2017_SI}%
  \BibitemOpen
  \bibfield  {author} {\bibinfo {author} {\bibfnamefont {J.}~\bibnamefont
  {Choi}}, \bibinfo {author} {\bibfnamefont {S.}~\bibnamefont {Choi}}, \bibinfo
  {author} {\bibfnamefont {G.}~\bibnamefont {Kucsko}}, \bibinfo {author}
  {\bibfnamefont {P.~C.}\ \bibnamefont {Maurer}}, \bibinfo {author}
  {\bibfnamefont {B.~J.}\ \bibnamefont {Shields}}, \bibinfo {author}
  {\bibfnamefont {H.}~\bibnamefont {Sumiya}}, \bibinfo {author} {\bibfnamefont
  {S.}~\bibnamefont {Onoda}}, \bibinfo {author} {\bibfnamefont
  {J.}~\bibnamefont {Isoya}}, \bibinfo {author} {\bibfnamefont
  {E.}~\bibnamefont {Demler}}, \bibinfo {author} {\bibfnamefont
  {F.}~\bibnamefont {Jelezko}}, \bibinfo {author} {\bibfnamefont {N.~Y.}\
  \bibnamefont {Yao}}, \ and\ \bibinfo {author} {\bibfnamefont {M.~D.}\
  \bibnamefont {Lukin}},\ }\href {\doibase 10.1103/PhysRevLett.118.093601}
  {\bibfield  {journal} {\bibinfo  {journal} {Phys. Rev. Lett.}\ }\textbf
  {\bibinfo {volume} {118}},\ \bibinfo {pages} {093601} (\bibinfo {year}
  {2017})},\ \bibinfo {note} {number: 9}\BibitemShut {NoStop}%
\bibitem [{\citenamefont {Akhmedzhanov}\ \emph {et~al.}(2017)\citenamefont
  {Akhmedzhanov}, \citenamefont {Gushchin}, \citenamefont {Nizov},
  \citenamefont {Nizov}, \citenamefont {Sobgayda}, \citenamefont {Zelensky},\
  and\ \citenamefont {Hemmer}}]{akhmedzhanov_microwave-free_2017_SI}%
  \BibitemOpen
  \bibfield  {author} {\bibinfo {author} {\bibfnamefont {R.}~\bibnamefont
  {Akhmedzhanov}}, \bibinfo {author} {\bibfnamefont {L.}~\bibnamefont
  {Gushchin}}, \bibinfo {author} {\bibfnamefont {N.}~\bibnamefont {Nizov}},
  \bibinfo {author} {\bibfnamefont {V.}~\bibnamefont {Nizov}}, \bibinfo
  {author} {\bibfnamefont {D.}~\bibnamefont {Sobgayda}}, \bibinfo {author}
  {\bibfnamefont {I.}~\bibnamefont {Zelensky}}, \ and\ \bibinfo {author}
  {\bibfnamefont {P.}~\bibnamefont {Hemmer}},\ }\href {\doibase
  10.1103/PhysRevA.96.013806} {\bibfield  {journal} {\bibinfo  {journal} {Phys.
  Rev. A}\ }\textbf {\bibinfo {volume} {96}},\ \bibinfo {pages} {013806}
  (\bibinfo {year} {2017})},\ \bibinfo {note} {number: 1}\BibitemShut {NoStop}%
\bibitem [{\citenamefont {Giri}\ \emph {et~al.}(2018)\citenamefont {Giri},
  \citenamefont {Gorrini}, \citenamefont {Dorigoni}, \citenamefont {Avalos},
  \citenamefont {Cazzanelli}, \citenamefont {Tambalo},\ and\ \citenamefont
  {Bifone}}]{giri_coupled_2018_SI}%
  \BibitemOpen
  \bibfield  {author} {\bibinfo {author} {\bibfnamefont {R.}~\bibnamefont
  {Giri}}, \bibinfo {author} {\bibfnamefont {F.}~\bibnamefont {Gorrini}},
  \bibinfo {author} {\bibfnamefont {C.}~\bibnamefont {Dorigoni}}, \bibinfo
  {author} {\bibfnamefont {C.~E.}\ \bibnamefont {Avalos}}, \bibinfo {author}
  {\bibfnamefont {M.}~\bibnamefont {Cazzanelli}}, \bibinfo {author}
  {\bibfnamefont {S.}~\bibnamefont {Tambalo}}, \ and\ \bibinfo {author}
  {\bibfnamefont {A.}~\bibnamefont {Bifone}},\ }\href {\doibase
  10.1103/PhysRevB.98.045401} {\bibfield  {journal} {\bibinfo  {journal} {Phys.
  Rev. B}\ }\textbf {\bibinfo {volume} {98}},\ \bibinfo {pages} {045401}
  (\bibinfo {year} {2018})},\ \bibinfo {note} {number: 4}\BibitemShut {NoStop}%
\bibitem [{\citenamefont {Hall}\ \emph {et~al.}(2016)\citenamefont {Hall},
  \citenamefont {Kehayias}, \citenamefont {Simpson}, \citenamefont {Jarmola},
  \citenamefont {Stacey}, \citenamefont {Budker},\ and\ \citenamefont
  {Hollenberg}}]{hall_detection_2016_SI}%
  \BibitemOpen
  \bibfield  {author} {\bibinfo {author} {\bibfnamefont {L.~T.}\ \bibnamefont
  {Hall}}, \bibinfo {author} {\bibfnamefont {P.}~\bibnamefont {Kehayias}},
  \bibinfo {author} {\bibfnamefont {D.~A.}\ \bibnamefont {Simpson}}, \bibinfo
  {author} {\bibfnamefont {A.}~\bibnamefont {Jarmola}}, \bibinfo {author}
  {\bibfnamefont {A.}~\bibnamefont {Stacey}}, \bibinfo {author} {\bibfnamefont
  {D.}~\bibnamefont {Budker}}, \ and\ \bibinfo {author} {\bibfnamefont
  {L.~C.~L.}\ \bibnamefont {Hollenberg}},\ }\href {\doibase
  10.1038/ncomms10211} {\bibfield  {journal} {\bibinfo  {journal} {Nat Commun}\
  }\textbf {\bibinfo {volume} {7}},\ \bibinfo {pages} {10211} (\bibinfo {year}
  {2016})},\ \bibinfo {note} {number: 1}\BibitemShut {NoStop}%
\bibitem [{\citenamefont {Delord}\ \emph {et~al.}(2018)\citenamefont {Delord},
  \citenamefont {Huillery}, \citenamefont {Schwab}, \citenamefont {Nicolas},
  \citenamefont {Lecordier},\ and\ \citenamefont {H\'etet}}]{DelordPRL_SI}%
  \BibitemOpen
  \bibfield  {author} {\bibinfo {author} {\bibfnamefont {T.}~\bibnamefont
  {Delord}}, \bibinfo {author} {\bibfnamefont {P.}~\bibnamefont {Huillery}},
  \bibinfo {author} {\bibfnamefont {L.}~\bibnamefont {Schwab}}, \bibinfo
  {author} {\bibfnamefont {L.}~\bibnamefont {Nicolas}}, \bibinfo {author}
  {\bibfnamefont {L.}~\bibnamefont {Lecordier}}, \ and\ \bibinfo {author}
  {\bibfnamefont {G.}~\bibnamefont {H\'etet}},\ }\href@noop {} {\bibfield
  {journal} {\bibinfo  {journal} {Phys. Rev. Lett.}\ }\textbf {\bibinfo
  {volume} {121}},\ \bibinfo {pages} {053602} (\bibinfo {year}
  {2018})}\BibitemShut {NoStop}%
\bibitem [{\citenamefont {Delord}(2019)}]{DelordPhD_SI}%
  \BibitemOpen
  \bibfield  {author} {\bibinfo {author} {\bibfnamefont {T.}~\bibnamefont
  {Delord}},\ }\emph {\bibinfo {title} {Spin-mechanics with micro-particles
  levitating in a Paul trap}},\ \href@noop {} {Ph.D. thesis},\ \bibinfo
  {school} {{\'E}cole Normale Sup{\'e}rieure} (\bibinfo {year}
  {2019})\BibitemShut {NoStop}%
\bibitem [{\citenamefont {Johansson}\ \emph {et~al.}(2012)\citenamefont
  {Johansson}, \citenamefont {Nation},\ and\ \citenamefont {Nori}}]{qutip1_SI}%
  \BibitemOpen
  \bibfield  {author} {\bibinfo {author} {\bibfnamefont {J.~R.}\ \bibnamefont
  {Johansson}}, \bibinfo {author} {\bibfnamefont {P.~D.}\ \bibnamefont
  {Nation}}, \ and\ \bibinfo {author} {\bibfnamefont {F.}~\bibnamefont
  {Nori}},\ }\href@noop {} {\bibfield  {journal} {\bibinfo  {journal} {Computer
  Physics Communications}\ }\textbf {\bibinfo {volume} {183}},\ \bibinfo
  {pages} {1760} (\bibinfo {year} {2012})}\BibitemShut {NoStop}%
\bibitem [{\citenamefont {Johansson}()}]{qutip2_SI}%
  \BibitemOpen
  \bibfield  {author} {\bibinfo {author} {\bibfnamefont {J.}~\bibnamefont
  {Johansson}},\ }\href@noop {} {\bibfield  {journal} {\bibinfo  {journal}
  {Comp. Phys. Comm}\ }\textbf {\bibinfo {volume} {184}},\ \bibinfo {pages}
  {1234}}\BibitemShut {NoStop}%
\bibitem [{\citenamefont {van Oort}\ and\ \citenamefont
  {Glasbeek}(1989)}]{van_oort_cross-relaxation_1989_SI}%
  \BibitemOpen
  \bibfield  {author} {\bibinfo {author} {\bibfnamefont {E.}~\bibnamefont {van
  Oort}}\ and\ \bibinfo {author} {\bibfnamefont {M.}~\bibnamefont {Glasbeek}},\
  }\href {\doibase 10.1103/PhysRevB.40.6509} {\bibfield  {journal} {\bibinfo
  {journal} {Phys. Rev. B}\ }\textbf {\bibinfo {volume} {40}},\ \bibinfo
  {pages} {6509} (\bibinfo {year} {1989})},\ \bibinfo {note} {number:
  10}\BibitemShut {NoStop}%
  
  
  
\end{thebibliography}
\end{document}